\documentclass[aps,prd,onecolumn,11pt,tightenlines,showkeys,notitlepage,longbibliography,nofootinbib,amsmath,amssymb]{revtex4-2}
\usepackage{graphicx}
\usepackage{xcolor}
\usepackage{xurl}
\usepackage[colorlinks=true,linkcolor=blue!50!black,citecolor=blue!50!black,urlcolor=blue!50!black]{hyperref}

\hypersetup{pdftitle={Free-streaming distance bounds and Lyman-alpha forest mass limits for nonthermal dark matter},pdfauthor={Zixuan Xu, Zhongjun Yang, Jiao Zhang}}
\begin{document}
\title{Free-streaming distance bounds and Lyman-\texorpdfstring{\(\alpha\)}{alpha} forest mass limits for nonthermal dark matter}
\author{Zixuan Xu}
\email[Contact author: ]{zixuanxu2010@gmail.com}
\author{Zhongjun Yang}
\email[Contact author: ]{zj\_yang@lit.edu.cn}
\author{Jiao Zhang}
\affiliation{Department of Mathematics and Physics, Luoyang Institute of Science and Technology, Luoyang, Henan 471023, China}

%%%%%%%%%%%%%%%%%%%%%%%%%%%%%%%%%%%%%%%%%%%%%%%%%%%%%%%%
\begin{abstract}
We derive sharp bounds on the mean comoving free-streaming distance that collisionless dark matter travels from the early radiation-dominated era to matter-radiation equality.
For a prescribed radiation and matter background, the bounds depend only on the mean and mean square of the present momentum per unit mass.
For fermions of known abundance, the Pauli exclusion principle raises the lower bound and lowers the upper bound at fixed moments.
These bounds determine the distance-matched mass interval within which the mean distance can equal that of a thermal relic at its observational lower mass limit.
For nonthermal dark matter produced solely by decays of thermalized particles, 
we compare this interval with one-sided 95\% credible lower mass limits inferred from Lyman-\(\alpha\) absorption in quasar spectra.
This comparison tests whether matching the mean distance can reproduce these observational limits for any momentum distribution with the same two moments.
When all decay products other than dark matter are massless, 
a power-law gas temperature history gives \(7.4\)~keV for two-body decay (scalar or fermion) and \(4.8\)~keV (scalar) or \(5.0\)~keV (fermion) for three-body decay.
Separate temperatures in four redshift bins lower these limits to \(6.3\), \(4.2\), and \(4.3\)~keV, respectively.
With a prior uniform in inverse mass, 
the observational limit can lie outside the mass interval in both two-body and three-body decay under either gas temperature history.
In three-body decay, changing the mass prior can move the limit for a scalar inside the interval.
\end{abstract}
%%%%%%%%%%%%%%%%%%%%%%%%%%%%%%%%%%%%%%%%%%%%%%%%%%%%%%%%%%

\maketitle
\enlargethispage{8pt}

%%%%%%%%%%%%%%%%%%%%%%%%%%%%%%%%%%%%%%%%%%%%%%%%%%%%%%%%%%%
\section{Introduction}
\label{sec:intro}

Collisionless dark matter streams out of overdense regions and suppresses the growth of structure on small scales~\cite{Ma:1995ey,Bode:2000gq,Viel:2005qj}.
The Lyman-\(\alpha\) forest in quasar spectra probes these scales through its one-dimensional flux power spectrum (hereafter flux power)~\cite{Viel:2013fqw}.
Measurements of the flux power set an observational lower mass limit for a thermal relic that decoupled while relativistic~\cite{Viel:2005qj,Irsic:2017ixq}.
For nonthermal dark matter, 
the suppression of structure depends on the momentum distribution established during production outside thermal equilibrium~\cite{Merle:2015oja,Heeck:2017xbu,DEramo:2020gpr}.

Previous studies inferred lower mass limits for nonthermal dark matter by comparison with a thermal relic at its observational lower mass limit.
They matched the mean comoving free-streaming distance (hereafter mean distance)~\cite{DEramo:2020gpr} or the velocity dispersion~\cite{Ballesteros:2020adh,DEramo:2025jsb} of the two species.
For specific momentum distributions, the mean distance and the linear matter power spectrum can give different lower mass limits~\cite{Konig:2016dzg,DEramo:2020gpr}.
Decay calculations can give the first two moments of the present momentum per unit mass analytically~\cite{DEramo:2020gpr,DEramo:2025jsb}.
Knowing these moments leaves open how much the mean distance can vary among compatible momentum distributions,
and whether that variation can account for differences between distance-matched masses and observational lower mass limits inferred directly from the flux power.

We derive sharp bounds on the mean distance at matter-radiation equality for all momentum distributions with the two prescribed moments.
We assume production early in radiation domination and a prescribed expansion history with radiation and matter.
For fermions of known abundance, the Pauli exclusion principle restricts the occupation of low momentum states~\cite{Tremaine:1979we,Carena:2021bqm}.
This restriction raises the lower bound and lowers the upper bound on the mean distance at fixed moments, 
and narrows the fractional width of the distance interval.

We apply the distance bounds to a one-state scalar and a two-state fermion produced in two-body decay or three-body decay.
We use the same flux power measurements~\cite{Viel:2013fqw} and likelihood~\cite{Hooper:2022byl} 
to infer observational lower mass limits for the thermal relic and for each full production distribution.
The two moments determine a distance-matched mass interval that bounds the masses matching the mean distance of the thermal relic at its observational lower mass limit.
We test whether the observational lower mass limit for each full production distribution lies in this interval.
With a prior uniform in inverse mass, 
the observational lower mass limit can lie outside the distance-matched mass interval in both two-body decay and three-body decay under either gas temperature history.
In three-body decay, changing the mass prior can move the limit for the one-state scalar inside the interval.

Section~\ref{sec:bounds} derives the distance bounds and determines how tightly the two moments constrain the mean distance.
Section~\ref{sec:calibration} uses the measured flux power to set the reference distance.
Section~\ref{sec:decays} compares the distance-matched mass interval with the observational lower mass limit for two-body decay and three-body decay.
Section~\ref{sec:conclusions} summarizes our conclusions.
Appendix~\ref{app:distance-proof} collects the proofs of the distance bounds and related asymptotics, 
and Appendix~\ref{app:checks} examines the sensitivity of the lower mass limits to the effective reionization redshift, mass prior, template interpolation weights, and perturbation treatment.

%%%%%%%%%%%%%%%%%%%%%%%%%%%%%%%%%%%%%%%%%%%%%%%%%%%%%%%%%%%%%%%%
\section{Bounds on the mean distance}
\label{sec:bounds}

Consider a stable dark matter particle \(\chi\) of mass \(m_\chi\) that propagates without collisions in a homogeneous expanding universe.
Let \(a\) denote the scale factor, with \(a=1\) today.
A collisionless particle with physical momentum of magnitude \(p(a)\) has conserved comoving momentum \(p_0=a p(a)\)~\cite{Ma:1995ey,Lesgourgues:2011rh}.
We use the dimensionless ratio \(r=p_0/m_\chi\), in natural units, where the reduced Planck constant \(\hbar\), 
the speed of light \(c\), and the Boltzmann constant \(k_{\rm B}\) are set to unity.
We retain the speed of light \(c\) explicitly when expressing a distance in megaparsecs (Mpc).
The speed of the particle in units of \(c\) is \(r/\sqrt{a^2+r^2}\).
If free streaming begins at scale factor \(a_i\), the comoving distance traveled by scale factor \(A>a_i\) is~\cite{DEramo:2020gpr,Liu:2024pjg}
\begin{equation}
 K_{i,A}(r)=\int_{a_i}^{A} W(a)\frac{r}{\sqrt{a^2+r^2}}\,{\rm d}a,
 \qquad W(a)=\frac{c}{a^2H(a)} .
 \label{eq:kernel}
\end{equation}
Here \(H(a)=\dot a/a\) is the Hubble rate, where the dot denotes a derivative with respect to cosmic time.
The quantity \(W(a)\,{\rm d}a\) is the comoving distance that light travels as the scale factor increases by \({\rm d}a\).
A particle at rest has \(K_{i,A}(0)=0\).
For sufficiently early production, we set \(a_i=0\) and denote this distance kernel by \(K_A=K_{0,A}\).
If free streaming begins at a positive \(a_i\), extending the integral to zero adds at most the comoving light-travel distance before \(a_i\), \(\int_0^{a_i}W(a)\,{\rm d}a\).

Let \(\mathcal P\) be the number-weighted probability distribution of \(r\).
For a smooth isotropic distribution, let \(f(r)\) denote the mean occupation per internal state.
The probability density in \(r\) is then proportional to \(r^2 f(r)\)~\cite{Ma:1995ey,Lesgourgues:2011rh}.
For any integrable function \(F(r)\), we write \(\mathbb E_{\mathcal P}[F(r)]=\int_{[0,\infty)}F(r)\,{\rm d}\mathcal P(r)\), including possible point masses.
The mean distance for this momentum distribution is \(\bar\lambda_{i,A}=\mathbb E_{\mathcal P}[K_{i,A}(r)]\).
We denote the mean of \(r\) by \(\mu\) and the ratio of the second moment to the squared mean by \(B\),
\begin{equation}
 \mu=\mathbb E_{\mathcal P}[r]>0,\qquad
 B=\frac{\mathbb E_{\mathcal P}[r^2]}{\mu^2}\geq 1.
 \label{eq:moments}
\end{equation}
The variance of \(r\) is \((B-1)\mu^2\).

%%%%%%%%%%%%%%%%%%%%%%%%%%%%%%%%%%%%%%%%%%%%%%%%%%%%%%%%%%%%%%%%
\subsection{Bounds from the two moments}

We consider a prescribed expansion history in which the comoving light-travel distance \(\int_0^A W(a)\,{\rm d}a\) is finite and \(W(a)/a\) is nonincreasing.
Both conditions hold in any spatially flat Friedmann universe that begins in a radiation- or matter-dominated era and contains radiation, nonrelativistic matter, and a cosmological constant.
Indeed, for total pressure \(p_{\rm tot}\) and energy density \(\rho_{\rm tot}\) in natural units, 
the Friedmann and energy-conservation equations give~\cite{Ma:1995ey} \(\mathrm d\ln(W/a)/\mathrm d\ln a=3(p_{\rm tot}/\rho_{\rm tot}-1)/2\), 
so \(W(a)/a\) is nonincreasing whenever \(p_{\rm tot}\leq\rho_{\rm tot}\).
Meanwhile, early radiation or matter domination makes \(\int_0^A W(a)\,{\rm d}a\) finite near \(a=0\).

For all momentum distributions on \(r\geq 0\) with a prescribed mean \(\mu\) and second moment \(B\mu^2\), the sharp bounds are
\begin{equation}
 L(\mu,B)=\frac{K_A(B\mu)}{B}
 \ \leq\ \bar\lambda_{0,A}\ \leq\
 U(\mu)=K_A(\mu).
 \label{eq:unbounded}
\end{equation}
A momentum distribution attaining the lower bound has a fraction \(1/B\) of particles moving with present momentum per unit mass \(B\mu\) and the remainder at rest.
The upper bound equals the distance traveled by a particle with \(r=\mu\).
For \(B>1\), the upper bound is an unattained supremum.
Momentum distributions approach this upper bound when a vanishing fraction of particles at large momentum carries the variance.
If \(B=1\), all particles have \(r=\mu\) and the two bounds coincide.
These bounds are sharp because each bound is attained or can be approached arbitrarily closely.
Since \(K_A(r)/r\) is convex (Appendix~\ref{app:two-moment-proof}), the moment inequality of Ref.~\cite{GPP1998} gives the lower bound.
Jensen's inequality~\cite{Jensen1906} for the concave kernel \(K_A(r)\) gives the upper bound.

The gap \(U-L\) gives the maximum amount by which \(U=K_A(\mu)\) can exceed the mean distance \(\bar\lambda_{0,A}\), 
and we call \(w=(U-L)/U\) the fractional width of the distance interval.
For propagation from \(a=0\) in an initially radiation-dominated background, write \(W_0=\lim_{a\to0}W(a)>0\).
When \(\int_0^A |W(a)-W_0|\,{\rm d}a/a\) is finite, the small-\(r\) expansion of \(K_A\) defines an effective scale factor,
\begin{equation}
 A_{\rm eff}=A\exp\!\left[\int_0^A\left(\frac{W(a)}{W_0}-1\right)\frac{{\rm d}a}{a}\right].
 \label{eq:scale-definition}
\end{equation}
At fixed \(A\) and \(B>1\), Appendix~\ref{app:cold} gives the cold limit of the fractional width,
\begin{equation}
 w \sim \frac{\ln B}{\ln(2A_{\rm eff}/\mu)}
 \qquad (\mu\to0).
 \label{eq:cold-fraction}
\end{equation}

Even at small \(\mu\), the fractional width decreases only logarithmically, 
and a larger second moment at the same \(\mu\) allows a larger fractional width.

%%%%%%%%%%%%%%%%%%%%%%%%%%%%%%%%%%%%%%%%%%%%%%%%%%%%%%%%%%%%%%%%
\subsection{Fermions of known abundance}
\label{sec:pauli}

The momentum distribution that minimizes the mean distance in Eq.~\eqref{eq:unbounded} concentrates a finite fraction of the relic at zero momentum for \(B>1\).
For fermions, the Pauli exclusion principle (hereafter Pauli exclusion) limits the occupation of each quantum state to one, 
so accommodating a nonzero fraction of the population requires a region of nonzero momentum-space volume~\cite{Tremaine:1979we}.
Let \(g_\chi\) count the populated internal states and \(n_0\) be the present number density, fixed by the rest-mass density \(m_\chi n_0\).
We call the sphere of lowest momentum states occupied by one fermion per state the filled ball.
At this number density the filled ball has radius \(m_\chi r_F\) in momentum space.
In natural units,
\begin{equation}
 r_F=\frac{1}{m_\chi}\left(\frac{6\pi^2n_0}{g_\chi}\right)^{1/3},
 \qquad {\rm d}\nu=\frac{3r^2}{r_F^3}\,{\rm d}r,\qquad
 {\rm d}\mathcal P=f(r)\,{\rm d}\nu,\quad 0\leq f(r)\leq1 .
 \label{eq:pauli-measure}
\end{equation}
The occupation satisfies \(\int f\,{\rm d}\nu=1\).
The measure \(\nu\) gives an upper bound on the fraction of the total population that each momentum interval can hold.
At fixed \(r_F\), the filled ball minimizes \(\mu\)~\cite{Carena:2021bqm}.
Integrating \(r\) over the filled ball in Eq.~\eqref{eq:pauli-measure} gives \(\mu\geq 3r_F/4\).

At fixed dark matter rest-mass density, decreasing \(m_\chi\) increases the number density and hence the momentum-space volume that the particles must occupy.
Equation~\eqref{eq:pauli-measure} gives \(r_F\propto m_\chi^{-4/3}\) for fixed \(g_\chi\).
When neither moment is prescribed, the filled ball gives the shortest mean distance because \(K_A(r)\) increases with \(r\).
With \(y=r/r_F\), this mean distance is
\begin{equation}
 \bar\lambda_{\rm ball}(m_\chi)=3\int_0^1 y^2 K_A\!\left(r_F(m_\chi)y\right)\,{\rm d}y .
 \label{eq:fermion-ball-distance}
\end{equation}

For each mean above the minimum, a single shell filled to unit occupation has the required abundance and mean.
This single shell gives the smallest possible second moment.
If the prescribed second moment is larger, the shortest mean distance comes from filling a central ball and a separate outer shell.
Their union is \(S_-=[0,r_1]\cup[r_2,r_3]\).
The single shell \(S_+=[r_-,r_+]\) determines the upper bound.
The radii follow from the moments of the filled regions,
\begin{align}
 \int_{S_-}(1,r,r^2)\,{\rm d}\nu&=(1,\mu,B\mu^2), &
 \int_{S_+}(1,r)\,{\rm d}\nu&=(1,\mu).
 \label{eq:pauli-radii}
\end{align}
These occupations give the sharp bounds for fermions of known abundance (Appendix~\ref{app:pauli}),
\begin{equation}
 L_{\rm P}=\int_{S_-}K_A(r)\,{\rm d}\nu
 \ \leq\ \bar\lambda_{0,A}\ \leq\
 U_{\rm P}=\int_{S_+}K_A(r)\,{\rm d}\nu .
 \label{eq:pauli-bounds}
\end{equation}
At fixed abundance, the two moments fix both the population of the outer shell in the minimizing distribution and the momenta that it occupies.
For the upper bound, the abundance and mean alone fix the single shell \(S_+\).
If the prescribed second moment exceeds that of \(S_+\), an arbitrarily small high-momentum population can carry the excess.
In that case, distributions with both moments approach \(U_{\rm P}\), while \(L_{\rm P}\) is attained.
When \(S_+\) has the prescribed second moment, it gives the only allowed distribution and the two bounds coincide.
At the smallest possible mean, \(\mu=3r_F/4\), the filled ball is the only possible occupation and has \(B=16/15\).

For any common initial scale factor \(0\leq a_i<A\) and positive integrable \(W\) on \([a_i,A]\), \(K_{i,A}\) is bounded and concave, 
so both upper bounds remain sharp (Appendix~\ref{app:distance-proof}).

%%%%%%%%%%%%%%%%%%%%%%%%%%%%%%%%%%%%%%%%%%%%%%%%%%%%%%%%%%%%%%%%
\subsection{Fractional width of the distance interval}
\label{sec:width}

The fractional width measures how tightly the two moments constrain the mean distance.
For fermions of known abundance we denote the fractional width by \(w_{\rm P}=(U_{\rm P}-L_{\rm P})/U_{\rm P}\).
If only the two moments are prescribed, monotonicity of \(K_A\) gives \(K_A(B\mu)\geq K_A(\mu)\).
Equation~\eqref{eq:unbounded} then gives \(w\leq 1-1/B\) for every expansion history satisfying the assumptions above.

For fermions, the occupied intervals in Eq.~\eqref{eq:pauli-radii} depend on the moments and the abundance, whereas the distance kernel also depends on the expansion history.
Write \(\widetilde S_\pm\) for the occupied intervals \(S_\pm\) expressed in \(y=r/r_F\).
The nonincreasing function \(W(a)/a\) can be written as the tail of a positive measure \(\mathcal M\), so that \(W(a)/a=\mathcal M([a,A])\) almost everywhere.
For each scale factor \(b\) in this representation, integrating \(ar/\sqrt{a^2+r^2}\) over \(a\) from zero to \(b\) gives \(G_b(r)=r(\sqrt{b^2+r^2}-r)\).
Exchanging the two positive integrals in Eq.~\eqref{eq:kernel} gives \(K_A(r)=\int_{(0,A]}G_b(r)\,{\rm d}\mathcal M(b)\).
In units of \(r_F\), let \(s=b/r_F\) and define the two integrals
\begin{equation}
 I_\pm(s)=\int_{\widetilde S_\pm} y\bigl(\sqrt{s^2+y^2}-y\bigr)\,3y^2\,{\rm d}y .
 \label{eq:width-layers}
\end{equation}
The ratio \(L_{\rm P}/U_{\rm P}\) is a weighted mean of \(I_-(s)/I_+(s)\), with positive weights proportional to \(I_+(b/r_F)\,{\rm d}\mathcal M(b)\).
It follows that
\begin{equation}
 w_{\rm P}\leq\sup_{s>0}\left[1-\frac{I_-(s)}{I_+(s)}\right].
 \label{eq:pauli-width-envelope}
\end{equation}
The right-hand side depends only on \(B\) and \(\mu/r_F\).

At fixed \(\mu/r_F\), a larger second moment can only lower the minimum mean distance.
An arbitrarily small population at sufficiently large \(r\) can supply the additional second moment at fixed mean.
At fixed \(B\), a larger \(\mu/r_F\) permits a larger occupation measure in units of \(r/\mu\), since the maximum probability in \({\rm d}(r/\mu)\) is \(3(\mu/r_F)^3(r/\mu)^2{\rm d}(r/\mu)\).
The supremum in Eq.~\eqref{eq:pauli-width-envelope} therefore cannot decrease along either direction within the allowed region.
Both integrals have leading term \(s^2/2\) at small \(s\) and \(s\mu/r_F\) at large \(s\), so \(1-I_-(s)/I_+(s)\) vanishes in both limits.

%%%%%%%%%%%%%%%%%%%%%%%%%%%%%%%%%%%%%%%%%%%%%%%%%%%%%%%%%%%%%%%%
\begin{figure}[tbp]
  \centering
  \includegraphics[width=.98\linewidth]{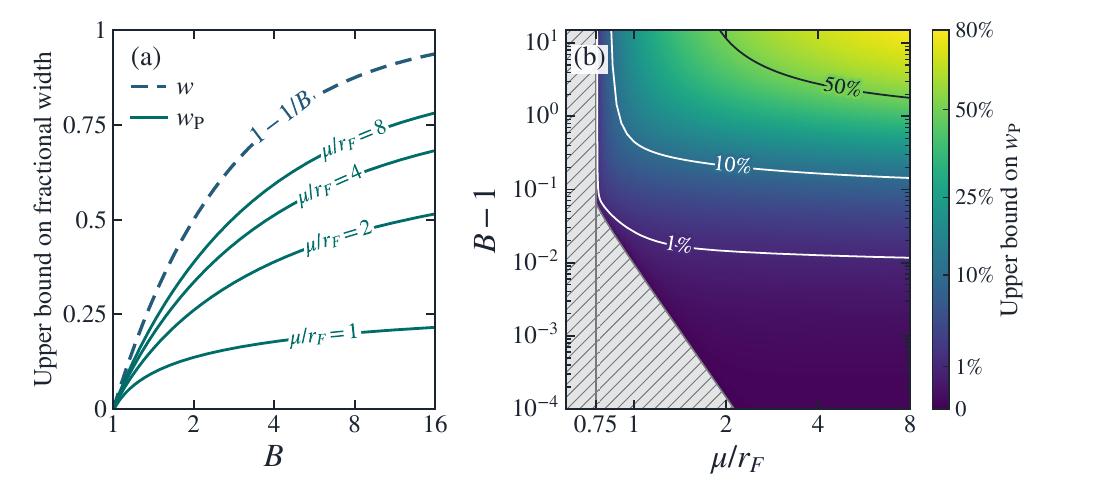}
\caption{Upper bounds on the fractional width of the distance interval.
In (a) the dashed curve bounds \(w=(U-L)/U\) when only the two moments are prescribed.
The solid curves bound \(w_{\rm P}=(U_{\rm P}-L_{\rm P})/U_{\rm P}\) for fermions of known abundance at the indicated values of \(\mu/r_F\).
Panel (b) shows the same upper bound on \(w_{\rm P}\) as a function of \(\mu/r_F\) and \(B-1\), the variance in units of \(\mu^2\).
Hatching marks moments incompatible with Pauli exclusion at the given abundance.
At the smallest allowed \(B\) for each \(\mu/r_F\), the fractional width vanishes.
At \(\mu/r_F=3/4\), only the filled ball with \(B=16/15\) is possible.}
  \label{fig:fermion-width-envelope}
\end{figure}
%%%%%%%%%%%%%%%%%%%%%%%%%%%%%%%%%%%%%%%%%%%%%%%%%%%%%%%%%%%%%%%%

The solid curves in Fig.~\ref{fig:fermion-width-envelope}(a) run from \(\mu/r_F=1\), near the minimum allowed mean, to \(\mu/r_F=8\), 
where Pauli exclusion allows a wider range of momentum distributions.
At each allowed pair \((\mu/r_F,B-1)\), the upper bound on \(w_{\rm P}\) in Fig.~\ref{fig:fermion-width-envelope}(b) applies to every expansion history considered in this section.
For \(\mu/r_F\leq1\) and \(B\leq2\), the mean \(\mu\) is at most \(r_F\) and the variance is at most \(\mu^2\).
Monotonicity and the value of the supremum at \(\mu/r_F=1\) and \(B=2\) give \(w_{\rm P}<13.64\%\) throughout the allowed part of this region (Appendix~\ref{app:pauli-width}).

%%%%%%%%%%%%%%%%%%%%%%%%%%%%%%%%%%%%%%%%%%%%%%%%%%%%%%%%%%%%%%%%
\section{Thermal calibration with Lyman-\texorpdfstring{\(\alpha\)}{alpha} forest data}
\label{sec:calibration}

%%%%%%%%%%%%%%%%%%%%%%%%%%%%%%%%%%%%%%%%%%%%%%%%%%%%%%%%%%%%%%%%
\subsection{Expansion history and flux power}

We adopted the Planck 2018 parameters for a spatially flat cosmology from the combined temperature, polarization, and lensing data~\cite{Planck:2018vyg}.
We write \(H_0=100h~{\rm km\,s^{-1}\,Mpc^{-1}}\) for the present Hubble rate and \(\omega_i=\Omega_i h^2\) for the physical density parameter of component \(i\), 
where \(\Omega_i\) is its present energy density in units of the critical density.
The present critical density is \(3H_0^2/(8\pi G)\) in natural units, where \(G\) is Newton's constant.
We used \(h=0.6736\), \(\omega_b=0.02237\) for baryons, and \(\omega_{\rm dm}=0.1200\) for dark matter.
The amplitude \(A_s\) and scalar spectral index \(n_s\) of the primordial curvature power spectrum satisfy \(\ln(10^{10}A_s)=3.044\) 
and \(n_s=0.9649\) at the comoving pivot wavenumber \(k=0.05~{\rm Mpc}^{-1}\).
The Thomson optical depth through reionization is \(\tau=0.0544\).
We retained the Planck baseline mass of \(0.06\)~eV for one active neutrino and the effective number of neutrinos \(3.046\)~\cite{Planck:2018vyg,Mangano:2005cc}.

Using the cold rest-mass density of dark matter gives a common expansion history for all momentum distributions.
Appendix~\ref{app:feedback} bounds the fractional decrease in the mean distance when the expansion includes the relic kinetic energy at fixed \(A\).
For neutrino pressure \(p_\nu\) and energy density \(\rho_\nu\), we assigned \(3p_\nu\) to radiation and \(\rho_\nu-3p_\nu\) to matter.
These assignments recover the radiation and matter contributions in the relativistic and nonrelativistic limits, respectively~\cite{Ma:1995ey,Lesgourgues:2011rh}.
With these assignments the matter and radiation densities are equal at \(A=a_{\rm eq}=2.938\times10^{-4}\), which defines matter-radiation equality for all of our distance comparisons.

The flux power measures fluctuations in the transmitted fraction of the quasar continuum relative to its mean.
The data \(D\) consist of 49 measurements of the flux power and their covariance at four redshifts \(z=4.2,4.6,5.0,5.4\), where \(1+z=1/a\).
The High Resolution Echelle Spectrometer (HIRES) on Keck~I and the Magellan Inamori Kyocera Echelle (MIKE) spectrograph supplied these measurements, which we call the HIRES/MIKE flux power~\cite{Viel:2013fqw}.
For each momentum distribution we computed the linear matter power spectrum \(P_{\rm m}(k,z)\) at comoving wavenumber \(k\) 
with the Cosmic Linear Anisotropy Solving System (CLASS)~\cite{Blas:2011rf,Lesgourgues:2011rh}.
In CLASS the relic supplies all of the dark matter and contributes its full energy density.
We assumed adiabatic primordial perturbations, so that the relic has no independent initial fluctuation relative to radiation 
on scales outside the horizon during production~\cite{Ma:1995ey,Strumia:2022qvj}.
We also computed the linear matter power spectrum for cold dark matter with the same cosmological parameters and primordial curvature power spectrum.
We define the linear transfer function as \(T(k)=[P_{\rm m}^{\chi}(k,0)/P_{\rm m}^{\rm cold}(k,0)]^{1/2}\).
The numerator and denominator are the linear total matter power spectra for the relic and cold dark matter models, respectively.

We used the Lyman-ABGD likelihood, which relates the linear transfer function to the flux power through a grid of hydrodynamical simulations~\cite{Hooper:2022byl}.
We fitted the linear transfer function with a weighted sum of simulated template transfer functions by bounded least squares.
At each redshift we applied the same weights to the suppression of the flux power in the simulations at the chosen gas parameters.

Thermal motions broaden hydrogen absorption lines, and gas pressure smooths fluctuations in the baryon density~\cite{Viel:2013fqw,Irsic:2017ixq}.
Both effects reduce the flux power on small scales, so the inferred dark matter mass depends on the gas temperature history.
For gas at the cosmic mean baryon density, let \(T_0(z)\) denote the temperature.
We considered a power-law gas temperature history, in which \(T_0(z)\) follows a power law in \(1+z\), and separate temperatures in four redshift bins~\cite{Irsic:2017ixq}.
For both gas temperature histories we used \(T_{\rm gas}=T_0(1+\delta_b)^{\gamma-1}\) for the local gas temperature, 
where \(\delta_b\) is the fractional deviation of the baryon density from its cosmic mean and \(\gamma(z)\) is the index of the temperature-density relation~\cite{Hui:1997dp,Irsic:2017ixq}.
We marginalized the Gaussian likelihood for the HIRES/MIKE flux power over the gas parameters, 
including the mean transmission in each redshift bin and the amplitude of fluctuations in the ultraviolet background~\cite{Esposito:2022plo,Hooper:2022byl}.

For the power-law gas temperature history we used \(T_0(z)=T_p[(1+z)/5.5]^{s_T}\) and \(\gamma(z)=\gamma_p[(1+z)/5.5]^{s_\gamma}\).
At the pivot redshift \(z=4.5\), within the range of the HIRES/MIKE data, \(T_p\) is the temperature at mean density and \(\gamma_p\) is the index of the temperature-density relation.
The exponents \(s_T\) and \(s_\gamma\) describe how these quantities vary with redshift.
We assigned uniform priors to \(T_p\) between 3000 and 20000~K, to \(\gamma_p\) between 0 and 2, and to each exponent between \(-5\) and 5~\cite{Esposito:2022plo}.
In all four redshift bins, we required temperatures covered by the grid of hydrodynamical simulations and \(1\leq\gamma(z)\leq1.7\).
For separate temperatures in four redshift bins, the joint prior is uniform over the supported part of \([3000,20000]^4\) in kelvin with a maximum difference of 5000~K between adjacent bins.

Both gas temperature histories use the same power law for \(\gamma(z)\) and a uniform prior on \([0,1]\) for the amplitude of fluctuations in the ultraviolet background.
In order of increasing redshift, the priors for the mean transmission are independent Gaussian distributions with means \((0.364,0.30,0.18,0.08)\) and standard deviation 0.04~\cite{Esposito:2022plo}.
We truncated each prior to the range covered by the simulations.

%%%%%%%%%%%%%%%%%%%%%%%%%%%%%%%%%%%%%%%%%%%%%%%%%%%%%%%%%%%%%%%%
\subsection{Reference thermal relic and distance criterion}

Let \(m\) denote the mass of the species in the fit.
We used a prior uniform in \(1/m\) for \(m\geq0.75\,{\rm keV}\), so that the cold dark matter limit corresponds to \(1/m=0\).
The lower endpoint specifies the mass range of our prior and lies below the lower mass limits reported here.
We obtain a posterior distribution for \(m\) at each choice of gas temperature history and production parameters.
Its fifth percentile gives the one-sided 95\% credible lower mass limit \(m_{95}\), so that \(\Pr(m<m_{95}\mid D)=0.05\).
The baseline fits use a prior uniform in inverse mass, bounded least squares, and a fixed effective reionization redshift \(z_{\rm reio}=7.69\).
This redshift sets the duration of pressure smoothing in the grid of hydrodynamical simulations~\cite{Viel:2013fqw,Hooper:2022byl}.
Within each gas temperature history we used the same mass and gas priors for every momentum distribution.

We used a thermal relic with two populated internal states to relate \(m_{95}\) to a mean distance.
For thermal relic mass \(m_{\rm th}\), the Fermi-Dirac occupation is \([\exp(p_0/T_{\rm th,0})+1]^{-1}\), 
where \(T_{\rm th,0}\) is the present temperature of the thermal relic in energy units.
Let \(T_{\gamma,0}=2.7255\)~K be the present photon temperature~\cite{Fixsen:2009ug} 
and \(T_{\nu,0}^{\rm id}=(4/11)^{1/3}T_{\gamma,0}\) the present temperature for instantaneous neutrino decoupling~\cite{Mangano:2005cc}.
Integrating the Fermi-Dirac occupation relates the present temperature to the thermal relic mass and abundance~\cite{Ballesteros:2020adh},
\begin{equation}
 T_{\rm th,0}=T_{\nu,0}^{\rm id}
 \left(\frac{94.1\,{\rm eV}\,\omega_{\rm dm}}{m_{\rm th}}\right)^{1/3}.
 \label{eq:thermal}
\end{equation}
The coefficient \(94.1\)~eV is the present critical density divided by \(h^2\) and by the number density of two ideally decoupled states at \(T_{\nu,0}^{\rm id}\).
For each gas temperature history, the reference thermal relic is this species evaluated at its own \(m_{95}\).
Its mean distance at matter-radiation equality defines the reference distance \(d_*\).

The distance criterion requires \(\bar\lambda_{0,A}\leq d_*\).
If \(U\leq d_*\), every momentum distribution with the two prescribed moments satisfies the distance criterion.
If \(L>d_*\), every such momentum distribution exceeds the reference distance.
If \(L\leq d_*<U\), the minimizing momentum distribution satisfies the distance criterion and momentum distributions sufficiently close to \(U\) exceed \(d_*\).
For a two-state fermion with the measured abundance the corresponding comparisons use \(L_{\rm P}\) and \(U_{\rm P}\).
To compare both gas temperature histories on the same scale, define \(\mu_*\) by \(K_A(\mu_*)=d_*\).
A particle with \(r=\mu_*\) therefore travels exactly the reference distance.

%%%%%%%%%%%%%%%%%%%%%%%%%%%%%%%%%%%%%%%%%%%%%%%%%%%%%%%%%%%%%%%%
\begin{figure}[tbp]
  \centering
  \includegraphics[width=.98\linewidth]{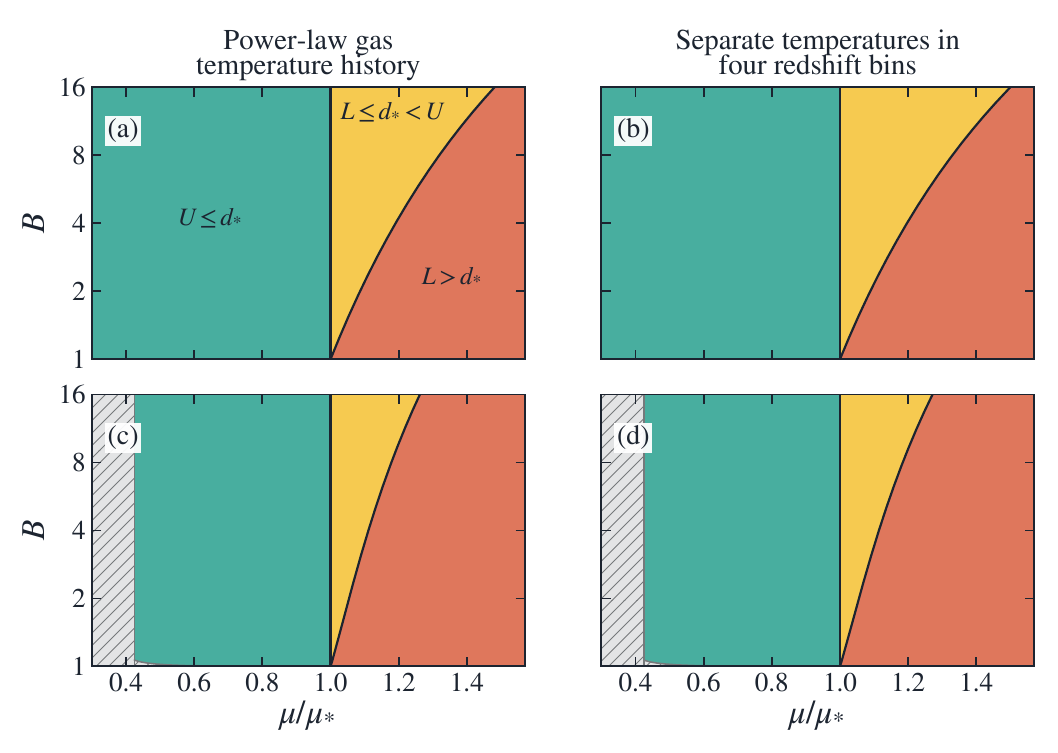}
\caption{Distance criterion at matter-radiation equality.
The inequalities in (a) label the colors in all four panels.
For \(U\leq d_*\), every momentum distribution with the two prescribed moments satisfies the distance criterion.
For \(L\leq d_*<U\), only some do, and for \(L>d_*\), none do.
Each horizontal axis uses \(\mu_*\) defined by \(K_A(\mu_*)=d_*\).
The upper panels fix only \(\mu\) and \(B\mu^2\).
The lower panels also impose the measured abundance \(\omega_{\rm dm}=0.1200\) and Pauli exclusion for a two-state fermion at the reference thermal relic mass, 
so that the inequalities use \(L_{\rm P}\) and \(U_{\rm P}\).
Hatching marks moment pairs with no allowed fermion distribution.
The left column uses the power-law gas temperature history, with \(d_*=0.1067\)~Mpc and \(\mu_*=2.363\times10^{-8}\).
The right column uses separate temperatures in four redshift bins, with \(d_*=0.1288\)~Mpc and \(\mu_*=2.914\times10^{-8}\).}
  \label{fig:decision}
\end{figure}
%%%%%%%%%%%%%%%%%%%%%%%%%%%%%%%%%%%%%%%%%%%%%%%%%%%%%%%%%%%%%%%%

Since \(K_A\) is strictly concave and increasing, \(\mu_*\) lies below the mean \(\mu\) of the reference thermal relic.
Both gas temperature histories use the same expansion history for \(K_A\), but their different thermal relic mass limits give different reference distances.
For the power-law gas temperature history we obtained \(m_{95}=3.31\)~keV and \(d_*=0.1067\)~Mpc.
For separate temperatures in four redshift bins we obtained \(m_{95}=2.83\)~keV and \(d_*=0.1288\)~Mpc, 
because allowing the gas temperature to vary independently in each redshift bin increases its degeneracy with free-streaming suppression~\cite{Irsic:2017ixq}.
An independent analysis of the HIRES/MIKE flux power alone found \(4.1\)~keV for a power-law gas temperature history 
and \(3.8\)~keV for separate temperatures in four redshift bins~\cite{Irsic:2017ixq}.
The lower baseline values here reflect differences in the hydrodynamical simulation grid, 
likelihood interpolation, and nuisance priors~\cite{Hooper:2022byl,Esposito:2022plo}.

In Fig.~\ref{fig:decision}(a,b), every momentum distribution with the prescribed moments satisfies the distance criterion for \(\mu\leq\mu_*\).
For \(\mu>\mu_*\), the distance criterion can still be satisfied if enough particles occupy low momenta.
Pauli exclusion restricts this possibility and rules out moment pairs for which the available quantum states cannot accommodate the measured abundance [Fig.~\ref{fig:decision}(c,d)].

%%%%%%%%%%%%%%%%%%%%%%%%%%%%%%%%%%%%%%%%%%%%%%%%%%%%%%%%%%%%%%%%
\subsection{Precision of the distance estimate}

The bounds also give a distance estimate when the full momentum distribution is unknown.
For an estimate \(\widehat\lambda\), we measure the relative error by \(|\widehat\lambda/\bar\lambda_{0,A}-1|\).
The largest relative error occurs at one end of the interval or arbitrarily close to it.
Equating the relative errors at the two ends gives the distance estimate with the smallest worst-case relative error \(\epsilon_*\),
\begin{equation}
 \widehat\lambda=\frac{2LU}{L+U},\qquad
 \epsilon_*=\frac{U-L}{U+L}=\frac{w}{2-w}.
 \label{eq:point-error}
\end{equation}
For fermions of known abundance the same formula uses \(L_{\rm P}\), \(U_{\rm P}\), and \(w_{\rm P}\).

\begin{figure}[tbp]
  \centering
  \includegraphics[width=.98\linewidth]{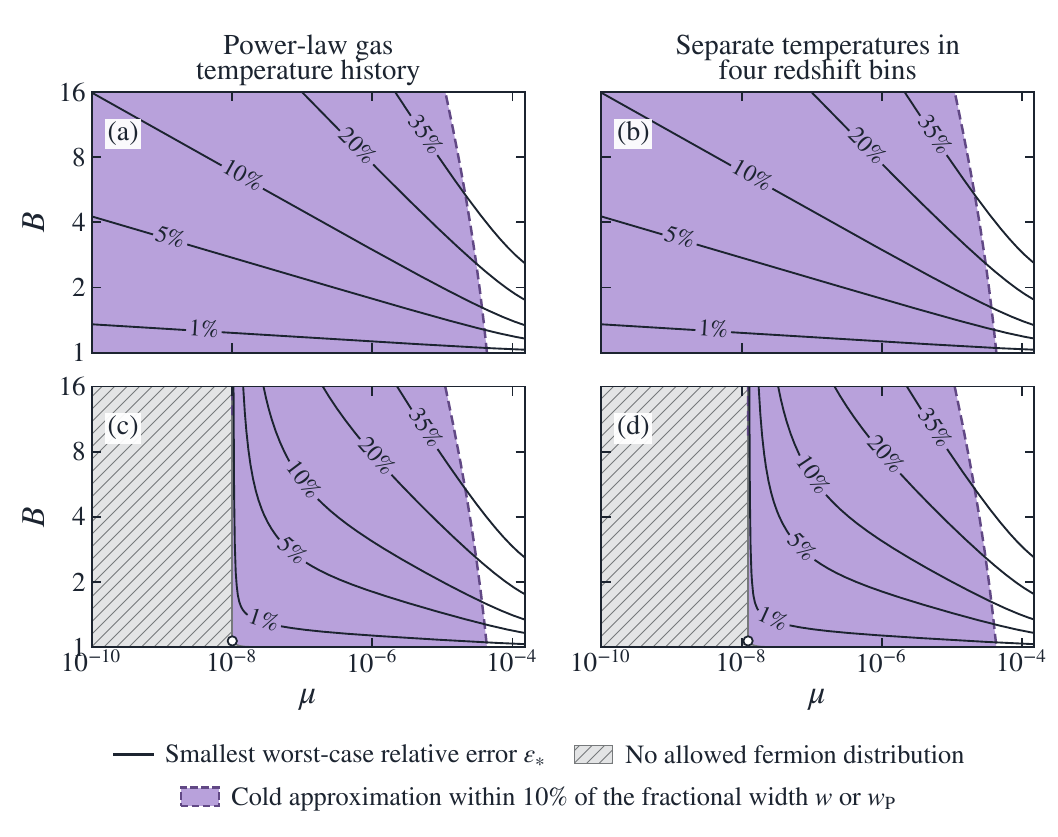}
\caption{Smallest worst-case relative error \(\epsilon_*\) of the mean distance estimate at matter-radiation equality.
Black contours give \(\epsilon_*\) for the two moments alone (top) and for a two-state fermion with \(\omega_{\rm dm}=0.1200\) and Pauli exclusion (bottom).
The fermion mass equals the reference thermal relic mass for the power-law gas temperature history (left, \(3.31\)~keV) or for separate temperatures in four redshift bins (right, \(2.83\)~keV).
Hatching marks moment pairs with no allowed fermion distribution.
The open circles mark the filled ball at \((\mu,B)=(3r_F/4,16/15)\), the only occupation at the smallest allowed \(\mu\).
The purple region satisfies \(\lvert w_{\rm cold}-w\rvert\leq0.1w\) for \(w>0\) (top) or \(\lvert w_{{\rm P},{\rm cold}}-w_{\rm P}\rvert\leq0.1w_{\rm P}\) for \(w_{\rm P}>0\) (bottom).}
  \label{fig:moment-precision}
\end{figure}

Figure~\ref{fig:moment-precision}(a,b) gives the same smallest worst-case relative error 
for both gas temperature histories because the upper panels depend only on \(\mu\), \(B\), and the prescribed expansion history.
The lower panels differ because the two reference thermal relic masses give different values of \(r_F\).
Where \(\epsilon_*\leq5\%\), the distance estimate in Eq.~\eqref{eq:point-error} has a relative error of at most \(5\%\) for every compatible momentum distribution.
Where \(\epsilon_*>5\%\), guaranteeing that precision requires more information about the momentum distribution.

We denote the cold approximation to the fractional width \(w\) in Eq.~\eqref{eq:cold-fraction} by \(w_{\rm cold}\).
For \(w_{\rm P}\) we obtain the cold approximation \(w_{{\rm P},{\rm cold}}\) by replacing \(K_A(r)\) in both integrals in Eq.~\eqref{eq:pauli-bounds} by \(W_0r\ln(2A_{\rm eff}/r)\).
Since both exact fractional widths can vanish, we compare the cold approximation with the corresponding exact fractional width only where that width is positive.
The plotted means range from \(\mu\ll a_{\rm eq}\), where a particle with \(r=\mu\) is nonrelativistic well before equality, to \(\mu\) of order \(a_{\rm eq}\), 
where the cold approximation can fail.
The purple shading marks the region where the cold approximation reproduces the fractional width to within a chosen relative tolerance of \(10\%\).

%%%%%%%%%%%%%%%%%%%%%%%%%%%%%%%%%%%%%%%%%%%%%%%%%%%%%%%%%%%%%%%%
%%%%%%%%%%%%%%%%%%%%%%%%%%%%%%%%%%%%%%%%%%%%%%%%%%%%%%%%%%%%%%%%

%%%%%%%%%%%%%%%%%%%%%%%%%%%%%%%%%%%%%%%%%%%%%%%%%%%%%%%%%%%%%%%%
\subsection{Effect of the fermion mass at fixed abundance}

\begin{figure}[tbp]
  \centering
  \includegraphics[width=.98\linewidth]{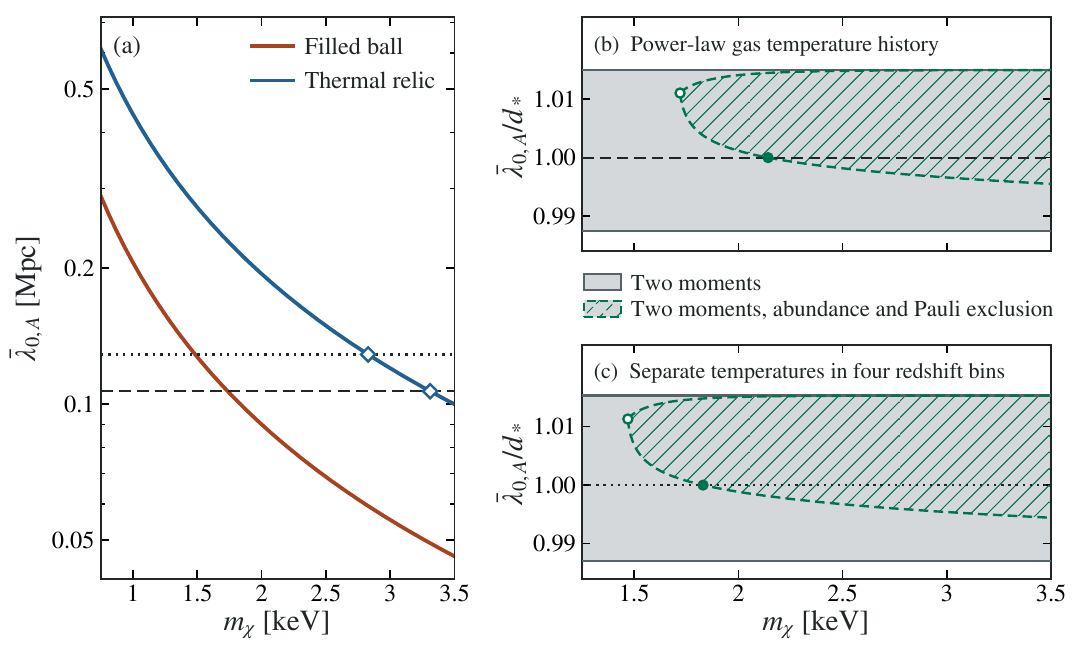}
\caption{Mean distance at matter-radiation equality for a two-state fermion with \(\omega_{\rm dm}=0.1200\).
In (a) the filled ball gives the shortest mean distance at each mass when only the abundance is prescribed.
Diamonds identify the two reference thermal relics on the thermal relic curve.
The filled ball reaches \(d_*=0.1067\)~Mpc for the power-law gas temperature history (dashed line) at \(1.74\)~keV and \(d_*=0.1288\)~Mpc 
for separate temperatures in four redshift bins (dotted line) at \(1.48\)~keV.
Panels (b) and (c) fix the two moments to those of the corresponding reference thermal relic, 
with \(B=1.303\) in both panels and \(\mu=2.402\times10^{-8}\) and \(2.964\times10^{-8}\), respectively.
The lower and upper edges of the gray band are \(L/d_*\) and \(U/d_*\).
Green hatching extends from \(L_{\rm P}/d_*\) to \(U_{\rm P}/d_*\) after imposing the measured abundance and Pauli exclusion.
Open circles mark the lower endpoints of the allowed mass ranges, \(1.72\)~keV in (b) and \(1.47\)~keV in (c).
Solid circles mark \(L_{\rm P}=d_*\) at \(2.14\)~keV in (b) and \(1.83\)~keV in (c).}
  \label{fig:fermion-mass-distance}
\end{figure}

Figure~\ref{fig:fermion-mass-distance}(a) shows how the mean distance of the filled ball in Eq.~\eqref{eq:fermion-ball-distance} decreases as \(m_\chi\) increases.
Its crossing of \(d_*\) marks the smallest mass for which the abundance alone permits the distance criterion.

%%%%%%%%%%%%%%%%%%%%%%%%%%%%%%%%%%%%%%%%%%%%%%%%%%%%%%%%%%%%%%%%
%%%%%%%%%%%%%%%%%%%%%%%%%%%%%%%%%%%%%%%%%%%%%%%%%%%%%%%%%%%%%%%%

Prescribing the two moments of the reference thermal relic introduces two distinct mass scales [Fig.~\ref{fig:fermion-mass-distance}(b,c)].
The open circle is the limiting mass at which \(\mu=3r_F/4\).
At that mass the filled ball is the only possible occupation and has \(B=16/15\), whereas the reference thermal relic has \(B=1.303\).
The prescribed moments are therefore possible only above the open circle.
Between the open and solid circles the moments are possible, yet the mean distance of every compatible momentum distribution exceeds \(d_*\).
The solid circle marks the first mass at which \(L_{\rm P}=d_*\).
Fixing the two moments of the reference thermal relic thus raises the smallest mass 
that permits the distance criterion from \(1.74\) to \(2.14\)~keV for the power-law gas temperature history [Fig.~\ref{fig:fermion-mass-distance}(a,b)].

%%%%%%%%%%%%%%%%%%%%%%%%%%%%%%%%%%%%%%%%%%%%%%%%%%%%%%%%%%%%%%%%
\section{Distance bounds and mass limits for decay-produced dark matter}
\label{sec:decays}

%%%%%%%%%%%%%%%%%%%%%%%%%%%%%%%%%%%%%%%%%%%%%%%%%%%%%%%%%%%%%%%%
\subsection{Production assumptions and the distance-matched mass interval}

We consider a bath particle \(B_1\) of mass \(M\) that decays into one stable dark matter particle \(\chi\) of mass \(m_\chi\) and a bath particle \(B_2\) of mass \(m_2\).
We call \(B_2\) and, in three-body decay, the additional massless bath particle \(B_3\) the bath daughters.
In both channels, \(\chi\) has one populated internal state for the one-state scalar and two for the two-state fermion.
The daughter-to-parent mass ratio is \(\rho=m_2/M\).
We considered \(0\leq\rho\leq0.90\) with \(m_\chi\ll M-m_2\).
At \(\rho=0\) all bath daughters are massless, while the largest mass ratio describes a small recoil without reaching the kinematic endpoint \(\rho=1\).

We assumed that radiation dominates during production, that the reheating temperature is well above \(M\), and that the dark sector initially contains no particles~\cite{DEramo:2020gpr}.
We took the bath particles to follow Maxwell-Boltzmann distributions with vanishing chemical potentials~\cite{Heeck:2017xbu,Decant:2021mhj}.
We used the ideal high-temperature Standard Model value \(106.75\) for both the effective energy and entropy degrees of freedom~\cite{Saikawa:2018rcs}.
The kinetic equation includes decay and inverse decay.
Particles already present in the final state increase the production rate for the one-state scalar (Bose enhancement) and reduce it for the two-state fermion (Pauli blocking)~\cite{Du:2021jcj}.
The decay rate fixes the abundance of \(\chi\) in an initially empty dark sector.
The resulting momentum distribution differs from its equilibrium form at the bath temperature, which makes \(\chi\) nonthermal~\cite{Du:2021jcj,DEramo:2020gpr}.
We call the momentum distribution obtained from this kinetic equation with the measured abundance the full production distribution.
If the same interaction also produces \(\chi\) through scattering of bath particles~\cite{Hall:2009bx}, 
the distance bounds apply to the moments of the combined momentum distribution, and the flux power must be evaluated for that distribution.

During production we write \(q=p/T\), where \(p\) is the magnitude of the physical momentum and \(T\) is the bath temperature in energy units.
Free propagation conserves \(q\) while the entropy degrees of freedom are constant.
After production the physical momentum redshifts in inverse proportion to the scale factor.
Today we write \(q=p_0/T_{*,0}\), where \(T_{*,0}\) is the bath temperature at production redshifted to the present.
Entropy conservation gives \(T_{*,0}=(3.91/106.75)^{1/3}T_{\gamma,0}\) with 3.91 and 106.75 for the effective entropy degrees of freedom today and at production, 
respectively~\cite{Heeck:2017xbu,Decant:2021mhj}.

Let \(E_\chi^*\) be the energy of \(\chi\) in the rest frame of \(B_1\), and define \(u=2E_\chi^*/M\) and \(\delta=1-\rho^2\).
In the limit \(m_\chi\ll M-m_2\), energy conservation fixes \(u=\delta\) for two-body decay~\cite{Heeck:2017xbu,DEramo:2020gpr}.
For three-body decay, let \(s_{23}=M^2(1-u)\) be the invariant mass squared of the two bath daughters.
The three-body phase-space measure factorizes into two successive two-body measures and an integral over \(s_{23}\)~\cite{DEramo:2020gpr,DEramo:2025jsb}.
The first two-body phase-space measure contributes \(1-s_{23}/M^2=u\).
The second two-body phase-space measure contributes \(1-m_2^2/s_{23}\) because \(B_3\) is massless.
A constant spin-averaged squared matrix element therefore gives the following probability density for \(u\), with normalization \(I_0(\rho)\),
\begin{equation}
 w_\rho(u)=\frac{u}{I_0(\rho)}\left(1-\frac{\rho^2}{1-u}\right),\qquad
 I_0(\rho)=\int_0^\delta u\left(1-\frac{\rho^2}{1-u}\right){\rm d}u,
 \qquad 0<u<\delta .
 \label{eq:recoil}
\end{equation}

In the dilute limit, the final-state occupation and inverse decay are negligible.
At fixed \(u\), averaging over the parent motion and the emission time gives \(q=uY\) for decays into empty states.
With \(\Gamma\) denoting the gamma function, \(Y\) has probability density \(g_Y(\ell)=\ell^{3/2}e^{-\ell}/\Gamma(5/2)\) for \(\ell>0\)~\cite{Heeck:2017xbu,Decant:2021mhj}.

In this limit, the number-weighted probability density for \(q\) in two-body decay is \(g_2(q)=g_Y(q/\delta)/\delta\), with mean \(5\delta/2\) and second moment \(35\delta^2/4\).
Averaging over the allowed recoil gives the three-body density
\begin{equation}
 g_3(q)=\int_0^\delta w_\rho(u)g_Y(q/u)\,\frac{{\rm d}u}{u}.
 \label{eq:threebody-moments}
\end{equation}

With \(i=2\) for two-body decay and \(i=3\) for three-body decay, write \(\phi(q)=g_i(q)/q^2\) for the momentum dependence of the dilute occupation.
In the absence of inverse decays and quantum statistical factors, decays would produce an occupation \(\mathcal N\phi(q)\), 
where the normalization \(\mathcal N\) is proportional to the decay rate.
To obtain the actual occupation \(f_\chi(q)\), let \(\eta=+1\) for the one-state scalar and \(\eta=-1\) for the two-state fermion.
For the assumed Maxwell-Boltzmann bath distributions, 
the gain and loss terms in the collision integral~\cite{Du:2021jcj} are proportional to \(1+\eta f_\chi\) and \(e^q f_\chi\), respectively.
Energy conservation and the vanishing bath chemical potentials give the two terms the same momentum dependence apart from these occupation factors.
Their difference supplies the factor \(1-f_\chi/f_{\rm eq}\), where \(f_{\rm eq}(q)=(e^q-\eta)^{-1}\) is the equilibrium occupation of \(\chi\) at the bath temperature.
Integrating at fixed \(q\) from an empty dark sector gives
\begin{equation}
 f_\chi(q)=f_{\rm eq}(q)\left[1-\exp\!\left(-\frac{\mathcal N\phi(q)}{f_{\rm eq}(q)}\right)\right].
 \label{eq:decay-occupation}
\end{equation}

At each \(m_\chi\) and \(\rho\), the measured abundance \(\omega_{\rm dm}\)~\cite{Planck:2018vyg} determines \(\mathcal N\).
Let \(\zeta\) denote the Riemann zeta function.
With \(g_\chi\) populated internal states, the convention of Eq.~\eqref{eq:thermal} gives
\begin{equation}
 \frac{g_\chi}{2}\,m_\chi
 \left(\frac{T_{*,0}}{T_{\nu,0}^{\rm id}}\right)^3
 \frac{\int_0^\infty q^2 f_\chi(q)\,{\rm d}q}{(3/2)\zeta(3)}
 =94.1\,{\rm eV}\,\omega_{\rm dm}.
 \label{eq:abundance}
\end{equation}
The denominator \((3/2)\zeta(3)\) is the Fermi-Dirac number integral per state for the thermal relic.
The number-weighted probability measure for \(q\) is \(q^2 f_\chi(q)\,{\rm d}q/\int_0^\infty q^2 f_\chi(q)\,{\rm d}q\).
With \(r=qT_{*,0}/m_\chi\), this measure gives both the mean distance and the moments that enter its bounds.

At fixed \(\rho\), we denote by \(m_{\rm dist}\) the mass at which the mean distance equals the reference distance \(d_*\) for a specified family of momentum distributions.
If only the two moments of the full production distribution are prescribed at each mass, the distance bounds give two mass thresholds.
The lower mass threshold \(m_-\) solves \(L=d_*\) and the upper mass threshold \(m_+\) solves \(U=d_*\).
For the two-state fermion we also calculate these thresholds with \(L_{\rm P}\) and \(U_{\rm P}\).
Along the model families considered here the mean distance decreases through each crossing as the mass increases.
Below \(m_-\) every compatible momentum distribution exceeds \(d_*\), while above \(m_+\) every compatible momentum distribution satisfies the distance criterion.
We call the closed interval \([m_-,m_+]\), including a limiting endpoint when the upper bound is a supremum, the distance-matched mass interval.
Integrating the full production distribution gives a particular \(m_{\rm dist}\) in this interval.

We inferred \(m_{95}\) separately from the HIRES/MIKE flux power using the full production distribution at each mass.
If \(m_{95}\) lies outside the distance-matched mass interval, 
changing the momentum distribution while preserving its two moments at each mass cannot make \(m_{\rm dist}\) equal to that \(m_{95}\).
Changing the momentum distribution can change the linear transfer function~\cite{Dienes:2021cxp} 
and hence the flux power predicted by hydrodynamical simulations~\cite{Murgia:2018now,Hooper:2022byl}.

%%%%%%%%%%%%%%%%%%%%%%%%%%%%%%%%%%%%%%%%%%%%%%%%%%%%%%%%%%%%%%%%
\subsection{Two-body decay}
\label{sec:twobody}

In the dilute limit, increasing \(\rho\) reduces the recoil momentum and rescales the momenta without changing \(B\).
At larger occupations the inverse decay and the final-state occupation in Eq.~\eqref{eq:decay-occupation} also change the shape of the full production distribution.

%%%%%%%%%%%%%%%%%%%%%%%%%%%%%%%%%%%%%%%%%%%%%%%%%%%%%%%%%%%%%%%%
\begin{figure}[tbp]
  \centering
  \includegraphics[width=.92\linewidth]{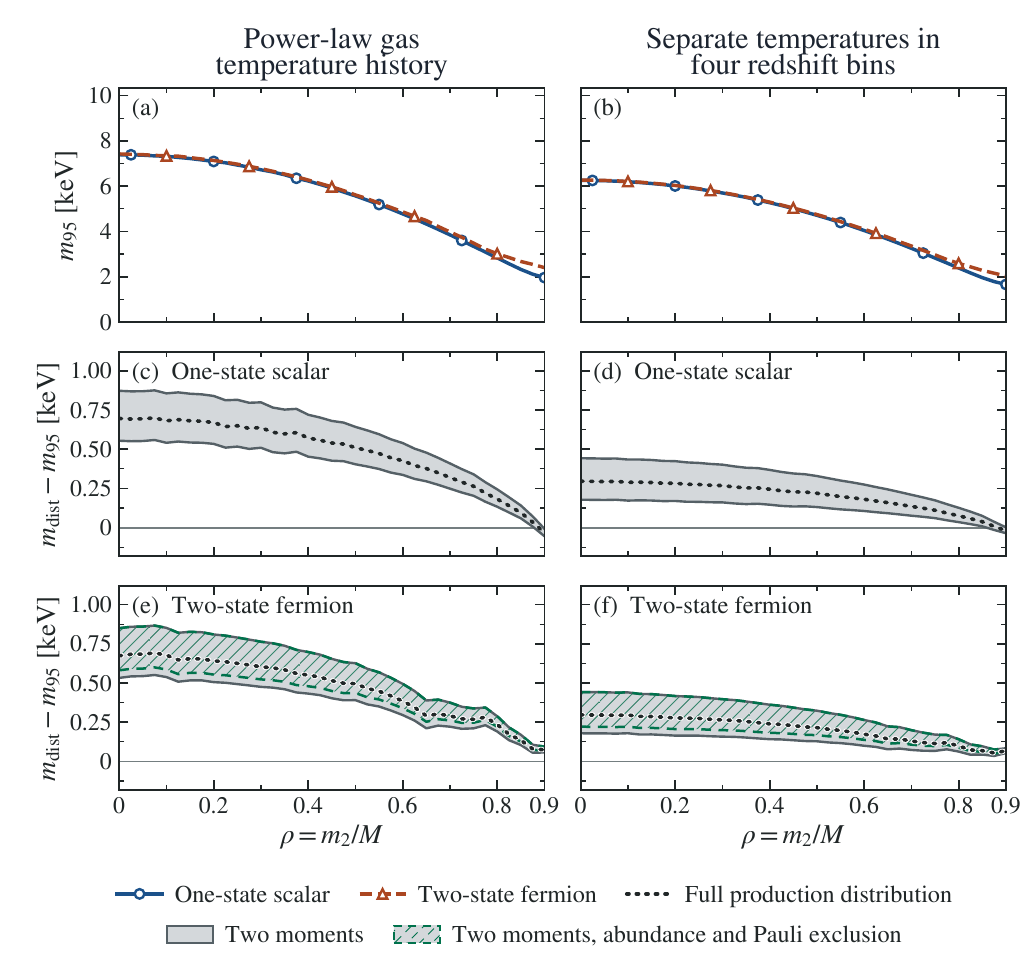}
\caption{Two-body decay.
Panels (a) and (b) show \(m_{95}\) from the HIRES/MIKE flux power~\cite{Viel:2013fqw} for the one-state scalar (circles) and the two-state fermion (triangles).
The horizontal axis gives the daughter-to-parent mass ratio \(\rho=m_2/M\).
The columns use the power-law gas temperature history (left) and separate temperatures in four redshift bins (right).
In the middle and bottom rows, the dotted curve gives \(m_{\rm dist}-m_{95}\) for the full production distribution.
The gray band extends from \(m_--m_{95}\) to \(m_+-m_{95}\) for the two moments alone.
Green hatching shows the interval after imposing the measured abundance and Pauli exclusion.
The solid gray and dashed green upper edges nearly coincide.
A band above zero means that every distance-matched mass in the interval exceeds \(m_{95}\).}
  \label{fig:two-body-decay-limits}
\end{figure}
%%%%%%%%%%%%%%%%%%%%%%%%%%%%%%%%%%%%%%%%%%%%%%%%%%%%%%%%%%%%%%%%

In Fig.~\ref{fig:two-body-decay-limits}(a,b), a heavier \(B_2\) allows a smaller \(m_\chi\) in the fits to the HIRES/MIKE flux power.
For the power-law gas temperature history, both particles give \(m_{95}\simeq7.4\)~keV at \(\rho=0\), 
while the one-state scalar gives a lower \(m_{95}\) than the two-state fermion at \(\rho=0.90\).
At large \(\rho\), smaller masses require larger phase-space occupations, 
so Bose enhancement concentrates the one-state scalar at lower momenta whereas Pauli blocking pushes the two-state fermion to higher momenta~\cite{Du:2021jcj}.
Separate temperatures in four redshift bins lower \(m_{95}\) for both particles.

For the one-state scalar [Fig.~\ref{fig:two-body-decay-limits}(c,d)], the power-law gas temperature history gives \(m_-=7.94\)~keV and \(m_{95}=7.39\)~keV at \(\rho=0\).
Even the shortest mean distance allowed by the two moments exceeds \(d_*\) for a range of masses above \(m_{95}\).
At \(\rho=0.90\), even \(m_+\) lies just below \(m_{95}\) in the baseline fit for the power-law gas temperature history [Fig.~\ref{fig:two-body-decay-limits}(c)].
With separate temperatures in four redshift bins, the baseline fit places \(m_{95}\) inside the distance-matched mass interval [Fig.~\ref{fig:two-body-decay-limits}(d)].
Near \(\rho=0.90\), the ordering also depends on the interpolation of the simulated flux power.

For the two-state fermion, \(m_->m_{95}\) already holds with the two moments alone [Fig.~\ref{fig:two-body-decay-limits}(e,f)].
This ordering holds at every sampled \(\rho\) for both gas temperature histories in the baseline fits.
Imposing the measured abundance and Pauli exclusion narrows a distance-matched mass interval that already lies above \(m_{95}\).

%%%%%%%%%%%%%%%%%%%%%%%%%%%%%%%%%%%%%%%%%%%%%%%%%%%%%%%%%%%%%%%%
\subsection{Three-body decay}
\label{sec:threebody}

For \(B_1\to B_2+B_3+\chi\), the invariant mass of \(B_2\) and \(B_3\) can vary at fixed \(\rho\).
The recoil momentum of \(\chi\) therefore spans a range even in the parent rest frame.
With the constant spin-averaged squared matrix element used here, the additional bath particle allows \(\chi\) to carry less momentum than in two-body decay at the same \(\rho\).

%%%%%%%%%%%%%%%%%%%%%%%%%%%%%%%%%%%%%%%%%%%%%%%%%%%%%%%%%%%%%%%%
\begin{figure}[tbp]
  \centering
  \includegraphics[width=.92\linewidth]{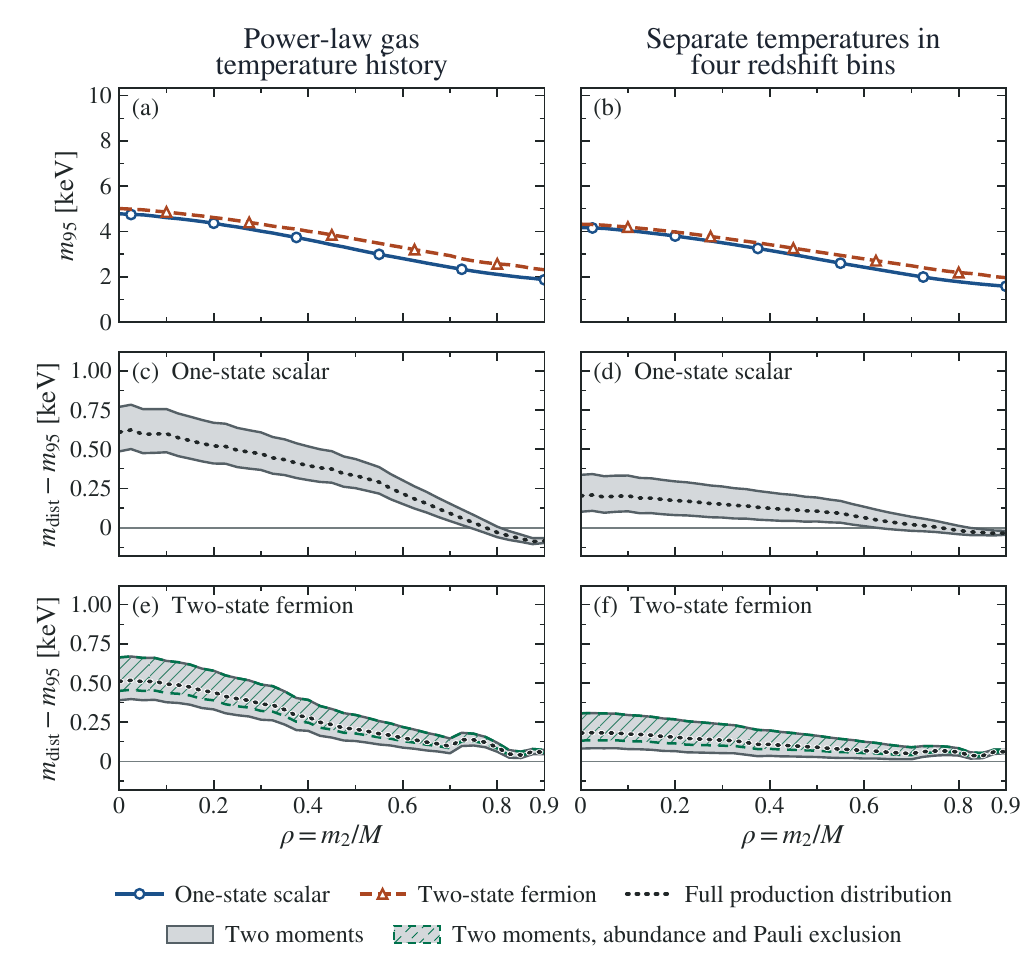}
\caption{Three-body decay.
Panels (a) and (b) show \(m_{95}\) from the HIRES/MIKE flux power~\cite{Viel:2013fqw} for the one-state scalar (circles) and the two-state fermion (triangles).
The horizontal axis gives the daughter-to-parent mass ratio \(\rho=m_2/M\).
The columns use the power-law gas temperature history (left) and separate temperatures in four redshift bins (right).
In the middle and bottom rows, the dotted curve gives \(m_{\rm dist}-m_{95}\) for the full production distribution.
The gray band extends from \(m_--m_{95}\) to \(m_+-m_{95}\) for the two moments alone.
Green hatching shows the interval after imposing the measured abundance and Pauli exclusion.
The solid gray and dashed green upper edges nearly coincide.
For the one-state scalar the entire band changes from positive to negative as \(\rho\) increases in the baseline fits.}
  \label{fig:three-body-decay-limits}
\end{figure}
%%%%%%%%%%%%%%%%%%%%%%%%%%%%%%%%%%%%%%%%%%%%%%%%%%%%%%%%%%%%%%%%

Figure~\ref{fig:three-body-decay-limits}(a,b) shows that at \(\rho=0\) both gas temperature histories give smaller \(m_{95}\) for three-body decay than for two-body decay.
For the power-law gas temperature history, the three-body limits are \(4.8\)~keV for the one-state scalar and \(5.0\)~keV for the two-state fermion.
In the baseline fits for both gas temperature histories, \(m_{95}\) decreases with increasing \(\rho\) over the sampled range for both the one-state scalar and the two-state fermion.

For the one-state scalar [Fig.~\ref{fig:three-body-decay-limits}(c,d)], \(m_->m_{95}\) at every sampled \(\rho\leq0.60\) for both gas temperature histories, 
whereas \(m_+<m_{95}\) at every sampled \(\rho\) from \(0.825\) to \(0.90\) in the baseline fits.
At large \(\rho\), Bose enhancement builds a pronounced low-momentum population alongside a high-momentum tail~\cite{Du:2021jcj}.
The low-momentum peak shortens the mean distance and lowers \(m_+\), 
while the high-momentum tail still suppresses the linear transfer function on the scales probed by the HIRES/MIKE flux power~\cite{Murgia:2018now,Dienes:2021cxp}.
At the baseline \(m_{95}\) inferred for the full production distribution, 
no momentum distribution with the prescribed two moments gives a mean distance equal to \(d_*\) in either sampled range.
With the baseline thermal calibration, changing the mass prior moves \(m_{95}\) inside the distance-matched mass interval 
at every sampled \(\rho\leq0.60\) for both gas temperature histories.
At the largest sampled mass ratios, \(m_{95}\) lies close to \(m_+\), 
and their ordering is sensitive to the mass prior and the numerical settings used for the thermal relic and the decay model.
Appendix~\ref{app:checks} examines how the mass prior, effective reionization redshift, and interpolation weights affect this ordering.

For the two-state fermion, \(m_->m_{95}\) already holds with the two moments alone [Fig.~\ref{fig:three-body-decay-limits}(e,f)].
This ordering holds at every sampled \(\rho\) for both gas temperature histories in the baseline fits.
At \(\rho=0.50\), imposing the measured abundance and Pauli exclusion raises \(m_-\) from \(3.79\) to \(3.84\)~keV for the power-law gas temperature history.
For masses between these two values of \(m_-\), the two moments alone permit the distance criterion, 
but every momentum distribution compatible with the measured abundance and Pauli exclusion gives a mean distance above \(d_*\).

%%%%%%%%%%%%%%%%%%%%%%%%%%%%%%%%%%%%%%%%%%%%%%%%%%%%%%%%%%%%%%%%

\section{Discussion and conclusions}
\label{sec:conclusions}

For early production of collisionless dark matter, 
the first two moments of the present momentum per unit mass give sharp bounds on the mean distance [Eq.~\eqref{eq:unbounded}].
The comoving light-travel distance must be finite from \(a=0\) to \(A\), and \(W(a)/a\) must be nonincreasing in the prescribed expansion history.
With early radiation domination and fixed \(B>1\), 
the fractional width of the distance interval decreases only logarithmically as \(\mu\) tends to zero [Eq.~\eqref{eq:cold-fraction}].
For fermions of known abundance, Pauli exclusion narrows the distance interval by limiting the occupation of low momentum states.
For every such expansion history, 
the fractional width \(w_{\rm P}\) is below \(13.64\%\) throughout the allowed region with \(\mu/r_F\leq1\) and \(B\leq2\).
The fractional width also fixes the smallest worst-case relative error of a distance estimate.

At its one-sided 95\% credible lower mass limit \(m_{95}\), a thermal relic travels a mean distance \(d_*\) by matter-radiation equality.
Comparing the distance bounds with this reference distance determines whether all, some, or none of the compatible momentum distributions satisfy the distance criterion (Fig.~\ref{fig:decision}).
For the two-state fermion at the measured abundance, some masses admit the prescribed moments even though the shortest mean distance exceeds \(d_*\) [Fig.~\ref{fig:fermion-mass-distance}(b,c)].

For nonthermal dark matter produced in decays, the bounds give the distance-matched mass interval.
We inferred \(m_{95}\) for each full production distribution from the same HIRES/MIKE flux power as for the reference thermal relic.
With massless bath daughters, separate temperatures in four redshift bins gave lower mass limits than the power-law gas temperature history for all four decay models.
In two-body decay, \(m_{95}\) fell from about \(7.4\) to \(6.3\)~keV for both the one-state scalar and the two-state fermion.
In three-body decay, \(m_{95}\) fell from \(4.8\) to \(4.2\)~keV for the one-state scalar and from \(5.0\) to \(4.3\)~keV for the two-state fermion.
In the baseline fits for both gas temperature histories, \(m_{95}\) decreased with increasing \(\rho\) for all four decay models over the sampled range.

For the one-state scalar in three-body decay, the entire distance-matched mass interval lay above \(m_{95}\) at small \(\rho\) in the baseline fits for both gas temperature histories.
At the largest sampled mass ratios, \(m_{95}\) was close to the upper mass threshold \(m_+\) [Fig.~\ref{fig:three-body-decay-limits}(c,d)].
For the two-state fermion, the same fits gave \(m_->m_{95}\) in both decay channels at every sampled \(\rho\) [Figs.~\ref{fig:two-body-decay-limits}(e,f) and~\ref{fig:three-body-decay-limits}(e,f)].

With the baseline thermal calibration, changing the mass prior moved \(m_{95}\) inside the distance-matched mass interval for the one-state scalar in three-body decay.
Both gas temperature histories gave this result at every sampled \(\rho\leq0.60\) (Appendix~\ref{app:checks}).
Our decay fits assume pure decays with a constant spin-averaged squared matrix element, whereas the distance bounds apply to any early-production momentum distribution.
Measurements of the flux power at smaller scales~\cite{Irsic:2023equ} could test whether \(m_{95}\) lies within the distance-matched mass interval at large \(\rho\) for the one-state scalar.
For nonthermal dark matter, matching the mean distance of a thermal relic~\cite{DEramo:2020gpr} cannot replace a flux-power fit whenever \(m_{95}\) lies outside the distance-matched mass interval.
Inferring reliable lower mass limits for nonthermal dark matter therefore requires the full linear transfer function, a flux-power likelihood, and an explicit mass prior.

%%%%%%%%%%%%%%%%%%%%%%%%%%%%%%%%%%%%%%%%%%%%%%%%%%%%%%%%%%%%%%%

%%%%%%%%%%%%%%%%%%%%%%%%%%%%%%%%%%%%%%%%%%%%%%%%%%%%%%%%%%%%%%
\begin{acknowledgments}
We used DeepSeek v4.1 to write code and edit the English prose.
The authors take full responsibility for the content of this work.
This work was supported by Natural Science Foundation of Henan (Grant No. 252300421996).
\end{acknowledgments}

\section*{Data Availability Statement}
The HIRES/MIKE flux-power measurements and the hydrodynamical interpolation grid are available 
with the public \href{https://github.com/brinckmann/montepython_public/tree/3.6/montepython/likelihoods/Lya_abgd}{Lyman-ABGD likelihood}~\cite{Viel:2013fqw,Hooper:2022byl}.
\section*{Code Availability Statement}
The code for the free-streaming calculation is available at \url{https://github.com/hunt4dm/dm-free-streaming-moment-bounds}.
The numerical calculations used NumPy~\cite{Harris:2020xlr}, SciPy~\cite{Virtanen:2019joe}, and CLASS~\cite{Blas:2011rf,Lesgourgues:2011rh}, 
and we produced the figures with Matplotlib~\cite{Hunter:2007ouj}.

%%%%%%%%%%%%%%%%%%%%%%%%%%%%%%%%%%%%%%%%%%%%%%%%%%%%%%%%%%%%%%%

%%%%%%%%%%%%%%%%%%%%%%%%%%%%%%%%%%%%%%%%%%%%%%%%%%%%%%%%%%%%%%
\appendix
%%%%%%%%%%%%%%%%%%%%%%%%%%%%%%%%%%%%%%%%%%%%%%%%%%%%%%%%%%%%%%%%
\section{Proofs of the distance bounds}
\label{app:distance-proof}

%%%%%%%%%%%%%%%%%%%%%%%%%%%%%%%%%%%%%%%%%%%%%%%%%%%%%%%%%%%%%%%%
\subsection{Two prescribed moments}
\label{app:two-moment-proof}

The lower bound in Eq.~\eqref{eq:unbounded} follows from the convexity of \(K_A(r)/r\) for \(r>0\).
To establish convexity, write \(V(a)=W(a)/a\).
Since \(V\) is nonincreasing, the positive measure \(\mathcal M\) introduced in Sec.~\ref{sec:width} satisfies \(V(a)=\int_{[a,A]}{\rm d}\mathcal M(b)\) almost everywhere.
The measure records the decreases of \(V\) and includes a mass \(V(A)\) at \(a=A\).
Interchanging the nonnegative integrals gives
\begin{equation}
 K_A(r)=\int_{(0,A]}G_b(r)\,{\rm d}\mathcal M(b),\qquad
 G_b(r)=\int_0^b\frac{ar}{\sqrt{a^2+r^2}}\,{\rm d}a
       =r\big(\sqrt{b^2+r^2}-r\big).
 \label{eq:layer-kernel}
\end{equation}
Each \(G_b(r)/r\) is strictly convex, so \(K_A(r)/r\) is strictly convex as well.
Under the probability measure \(r\,{\rm d}\mathcal P(r)/\mu\), the mean of \(r\) is \(B\mu\).
Jensen's inequality under this measure gives \(\mathbb E_{\mathcal P}[K_A(r)]\geq K_A(B\mu)/B\).
Equality requires \(r=B\mu\) for every moving particle.
The mean then fixes the fraction of moving particles to \(1/B\).

For the upper bound, the speed in Eq.~\eqref{eq:kernel} is strictly concave in \(r\geq0\) for each \(a>0\).
Jensen's inequality under \(\mathcal P\) therefore gives \(\mathbb E_{\mathcal P}[K_A(r)]\leq K_A(\mu)\).
To prove sharpness when \(B>1\), let \(v=(B-1)\mu^2\) and take any \(R>B\mu\).
Put probability \(v/[v+(R-\mu)^2]\) at \(R\) and the rest at \(\mu-v/(R-\mu)\).
This distribution has the prescribed mean and variance.
As \(R\to\infty\), the lower momentum tends to \(\mu\) and the probability at \(R\) vanishes.
The bounded, continuous kernel then gives a mean distance that approaches \(K_A(\mu)\).
Strict concavity excludes equality at positive variance.

%%%%%%%%%%%%%%%%%%%%%%%%%%%%%%%%%%%%%%%%%%%%%%%%%%%%%%%%%%%%%%%%
\subsection{Fermions of known abundance}
\label{app:pauli}

For a fermion, Pauli exclusion limits the occupation of each state to \(0\leq f\leq1\) in Eq.~\eqref{eq:pauli-measure}.
Moment problems with an upper bound on the density admit extrema with fully occupied intervals~\cite{McFadden1975,Cerf1994}.
We first determine which moments are possible on the unbounded momentum domain.
Filling the lowest states, \(0\leq r\leq r_F\), minimizes the mean at fixed abundance.
This filled ball is the only occupation with \(\mu=3r_F/4\), and its second moment is \(3r_F^2/5\).

For each \(\mu>3r_F/4\), the abundance and mean uniquely determine the single shell \(S_+=[r_-,r_+]\) filled to unit occupation.
Normalizing this single shell gives \(r_+^3-r_-^3=r_F^3\), and its mean increases continuously from \(3r_F/4\) to infinity as \(r_-\) increases.
Let \(\mathbf{1}_S\) denote the function that is one on \(S\) and zero elsewhere.
For any other allowed occupation \(f\), the product
\[
 (r-r_-)(r-r_+)\big[f(r)-\mathbf{1}_{S_+}(r)\big]
\]
is nonnegative everywhere.
The two occupations have the same number and mean, so the integral of the product against \({\rm d}\nu\) equals the difference between their second moments.
Hence \(S_+\) gives the smallest possible second moment, \(M_{2,\min}=\int_{S_+}r^2\,{\rm d}\nu\).
Equality fixes the occupation up to sets of zero measure.

For each larger second moment, Eq.~\eqref{eq:pauli-radii} specifies a unique central ball and outer shell, \(S_-=[0,r_1]\cup[r_2,r_3]\).
Let \(t=(r_1/r_F)^3\) be the fraction of particles in the central ball.
At fixed \(0\leq t<1\), moving the outer shell outward at fixed population \(1-t\) raises the mean continuously from \(3r_F/4\) to infinity.
The prescribed mean fixes \(r_2\) and \(r_3\) at each \(t\).
Differentiating the number and mean constraints gives
\begin{equation}
 \frac{{\rm d}}{{\rm d}t}\int_{S_-}r^2\,{\rm d}\nu
  =(r_2-r_1)(r_3-r_1)>0.
\end{equation}
At \(t=0\), the second moment equals \(M_{2,\min}\).
As \(t\to1\), the vanishing outer population still contributes \(\mu-3r_F/4>0\) to the mean.
Its second moment must therefore diverge.
The two regions together realize each prescribed second moment above \(M_{2,\min}\).

%%%%%%%%%%%%%%%%%%%%%%%%%%%%%%%%%%%%%%%%%%%%%%%%%%%%%%%%%%%%%%%%
\begin{figure}[tbp]
  \centering
  \includegraphics[width=.98\linewidth]{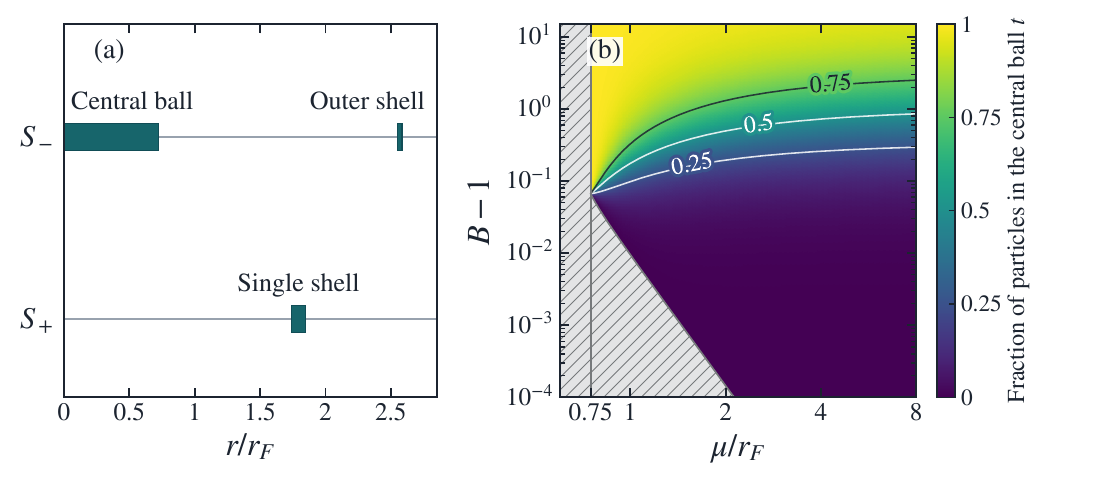}
\caption{Fermion occupations that determine the distance bounds.
Panel (a) fixes the abundance and the two moments to those of a thermal relic with two populated internal states.
Colored intervals have unit occupation.
The central ball and outer shell \(S_-\) attain \(L_{\rm P}\).
The single shell \(S_+\) has the prescribed abundance and mean, but has a smaller second moment.
Its mean distance is \(U_{\rm P}\), which momentum distributions with both prescribed moments can approach.
Panel (b) gives the fraction of particles in the central ball, \(t=(r_1/r_F)^3\), as a function of \(\mu/r_F\) and \(B-1\).
Hatching marks moments incompatible with Pauli exclusion.
The thermal relic in (a) has \(\mu/r_F\simeq1.80\) and \(B-1\simeq0.30\).
For this abundance and these moments, the central ball contains about \(38\%\) of the particles in the minimizing occupation.}
  \label{fig:occupations}
\end{figure}
%%%%%%%%%%%%%%%%%%%%%%%%%%%%%%%%%%%%%%%%%%%%%%%%%%%%%%%%%%%%%%%%

To prove that \(S_-\) gives the lower bound on the mean distance, we use the fact that \(G_b'''(r)=3b^4/(b^2+r^2)^{5/2}>0\).
On compact subsets of \(r>0\), a constant times \(b^2\) bounds this derivative, and \(\int b^2\,{\rm d}\mathcal M(b)=2\int_0^A W(a)\,{\rm d}a\) is finite.
Equation~\eqref{eq:layer-kernel} thus gives \(K_A'''(r)>0\).
If a quadratic \(Q\) interpolates \(K_A\) at \(r_1,r_2,r_3\), the interpolation remainder \(K_A(r)-Q(r)\) has the sign of \((r-r_1)(r-r_2)(r-r_3)\).
It is negative inside the two occupied regions and positive outside.
It follows that \((K_A-Q)(f-\mathbf{1}_{S_-})\geq0\).
The number and the first two moments fix the integral of \(Q\), so integration gives the lower bound in Eq.~\eqref{eq:pauli-bounds}.

For the upper bound, the secant line through the values of \(K_A\) at \(r_-\) and \(r_+\) lies below the kernel inside \(S_+\) and above it outside by strict concavity.
Subtracting this line and integrating against \((f-\mathbf{1}_{S_+})\,{\rm d}\nu\) proves that \(\int K_A f\,{\rm d}\nu\leq\int_{S_+}K_A\,{\rm d}\nu\).
Only the single shell \(S_+\) attains equality.
If \(B\mu^2>M_{2,\min}\), mix \(S_+\) with a central ball and outer shell of the same mean and a second moment \(M_2>B\mu^2\).
A fraction \((B\mu^2-M_{2,\min})/(M_2-M_{2,\min})\) of the latter gives the required second moment and preserves \(f\leq1\).
This fraction tends to zero as \(M_2\to\infty\).
Boundedness of \(K_A\) then gives convergence to the mean distance of \(S_+\).

Figure~\ref{fig:occupations}(a) shows that the central ball in \(S_-\) holds about \(38\%\) of the particles 
when the abundance and the two moments match those of a thermal relic with two populated internal states.
At fixed mean, a larger second moment puts a larger fraction of particles in the central ball and moves the outer shell to larger momentum [Fig.~\ref{fig:occupations}(b)].

%%%%%%%%%%%%%%%%%%%%%%%%%%%%%%%%%%%%%%%%%%%%%%%%%%%%%%%%%%%%%%%%

%%%%%%%%%%%%%%%%%%%%%%%%%%%%%%%%%%%%%%%%%%%%%%%%%%%%%%%%%%%%%%%%
\subsection{Upper bound on the fractional width for fermions}
\label{app:pauli-width}

To evaluate Eq.~\eqref{eq:pauli-width-envelope}, the occupation integrals need only be evaluated over the central ball and outer shell or over the single shell.
At either end of the positive \(s\) axis, the following bounds control the contribution to the supremum,
\begin{equation}
 0\leq 1-\frac{I_-(s)}{I_+(s)}
 \leq \min\!\left(\frac{3s^2}{4},\,\frac{B\mu}{s r_F}\right).
 \label{eq:width-tails}
\end{equation}
For the first bound, the factor \(y(\sqrt{s^2+y^2}-y)\) in Eq.~\eqref{eq:width-layers} divided by \(s^2\) is \(y/(\sqrt{s^2+y^2}+y)\).
It lies between \(1/2-s^2/(8y^2)\) and \(1/2\).
The filled ball maximizes the mean of \(y^{-2}\) at fixed abundance and gives a value of three.
For the second bound, the kernel lies between \(sy-y^2\) and \(sy\), whose integrals depend only on the prescribed moments.
On the remaining finite interval in \(\ln s\), we bounded the interpolation of \(\ln(I_-/I_+)\) using its second derivative.
The logarithmic derivative of the kernel with respect to \(s\) lies between one and two, and its derivative with respect to \(\ln s\) lies between \(-2/(3\sqrt{3})\) and zero.
For either occupation, differentiating the logarithm of its integral adds the variance of the first derivative, at most \(1/4\), to the weighted mean of the second derivative.
The absolute second derivative of \(\ln(I_-/I_+)\) with respect to \(\ln s\) is therefore at most \(1/4+2/(3\sqrt{3})\).
Together with Eq.~\eqref{eq:width-tails}, this bound encloses the supremum between sampled scales.
At \(\mu/r_F=1\) and \(B=2\), these bounds place the supremum in Eq.~\eqref{eq:pauli-width-envelope} below \(13.64\%\).

%%%%%%%%%%%%%%%%%%%%%%%%%%%%%%%%%%%%%%%%%%%%%%%%%%%%%%%%%%%%%%%%
\subsection{Cold limit during radiation domination}
\label{app:cold}

With the assumptions of Eq.~\eqref{eq:cold-fraction}, \(W(a)\to W_0>0\) as \(a\to0\) and \(\int_0^A|W(a)-W_0|\,{\rm d}a/a<\infty\).
Subtracting \(W_0\operatorname{arsinh}(A/r)\) from \(K_A(r)/r\) leaves the integral of \([W(a)-W_0]/\sqrt{a^2+r^2}\).
By dominated convergence, this integral tends to \(\int_0^A[W(a)-W_0]\,{\rm d}a/a\).
The expansion of the inverse hyperbolic sine then gives
\begin{equation}
 K_A(r)=W_0r\big[\ln(2A_{\rm eff}/r)+o(1)\big]\qquad(r\to0),
 \label{eq:effective-scale}
\end{equation}
with \(A_{\rm eff}\) as defined in Eq.~\eqref{eq:scale-definition}.
Evaluating this expression at \(\mu\) and \(B\mu\) with fixed \(B>1\) gives \(U-L=W_0\mu[\ln B+o(1)]\).
Division by \(U=K_A(\mu)\) gives the fractional width in Eq.~\eqref{eq:cold-fraction}.

%%%%%%%%%%%%%%%%%%%%%%%%%%%%%%%%%%%%%%%%%%%%%%%%%%%%%%%%%%%%%%%%
\subsection{The reference expansion}
\label{app:feedback}

The distance calculation treats the relic as cold when it contributes to \(H(a)\).
To bound the resulting change in mean distance, we hold its present rest-mass density \(\sigma=m_\chi n_0\), the other component densities, and the final scale factor \(A\) fixed.
The physical momenta of the collisionless particles redshift as \(a^{-1}\), so the relic energy density satisfies \(a^4\rho_\chi(a)=\sigma\mathbb E_{\mathcal P}[\sqrt{a^2+r^2}]\) in natural units~\cite{Ma:1995ey,Lesgourgues:2011rh}.
Since \(0\leq\sqrt{a^2+r^2}-a\leq r\), the relic increases \(a^4\rho_{\rm tot}\) by at most \(\sigma\mu\) relative to the cold reference cosmology.
In the cold reference cosmology, the early radiation density extrapolated to today is a lower bound on \(a^4\rho_{\rm tot}\).
We denote the physical density parameter of this radiation by \(\omega_{\rm rad}\).
The Friedmann equation then bounds the fractional decrease in the mean distance for the same \(\mathcal P\) by \(1-(1+\omega_{\rm dm}\mu/\omega_{\rm rad})^{-1/2}\).

%%%%%%%%%%%%%%%%%%%%%%%%%%%%%%%%%%%%%%%%%%%%%%%%%%%%%%%%%%%%%%%%
\section{Sensitivity of the lower mass limits}
\label{app:checks}

The effective reionization redshift in the grid of hydrodynamical simulations changes the time available for gas pressure to smooth density fluctuations~\cite{Viel:2013fqw,Hooper:2022byl}.
Holding the linear matter power spectrum fixed, we replace the fiducial value \(z_{\rm reio}=7.69\) with a uniform prior from 7 to 15, 
the range used to probe uncertainty in pressure smoothing~\cite{Esposito:2022plo}.
Because \(z_{\rm reio}=7.69\) lies near the lower end of this interval, marginalization admits earlier reionization, 
which lengthens gas pressure smoothing and leaves less room for dark matter free-streaming suppression.
Consequently, marginalization raises the one-sided 95\% credible lower mass limit 
for the thermal relic from 3.31 to 3.46~keV for the power-law gas temperature history and from 2.83 to 2.97~keV for separate temperatures in four redshift bins.
For all four decay models at \(\rho=0\) and 0.90, the lower mass limits increase by 0.05--0.46~keV across the two gas temperature histories.

The prior on the dark matter mass affects the comparison even when the likelihood for the HIRES/MIKE flux power and reference thermal relic stay fixed.
To compare two priors on the same mass range, we used a dimensionless variable based on the present root-mean-square momentum per unit mass,
\begin{equation}
 \xi=\left[\frac{\sqrt{\mathbb E_{\mathcal P}[r^2]}}
 {\sqrt{\mathbb E_{\rm th,1}[r^2]}}\right]^{3/4},
 \label{eq:speed-prior}
\end{equation}
where \(\mathbb E_{\rm th,1}\) denotes the number-weighted average for a thermal relic of mass 1~keV with the same abundance.
For the thermal family, Eq.~\eqref{eq:thermal} gives \(\xi=1\,{\rm keV}/m_{\rm th}\).
For all four decay models and both gas temperature histories, 
we compared priors uniform in \(1/m_\chi\) and in \(\xi\) at every sampled \(\rho\) on the same mass range, \(m_\chi\geq0.75~{\rm keV}\).
In each fit we integrated the posterior probability density for the mass up to the cold dark matter limit, \(m\to\infty\).
The two priors coincide for the thermal relic, so the reference distance and the distance-matched mass interval are the same in both fits.

In two-body decay, both gas temperature histories give \(m_{95}<m_-\) for the prior uniform in \(1/m_\chi\) and \(m_{95}>m_-\) for the prior uniform in \(\xi\) at every sampled \(\rho\leq0.60\).
Both inequalities hold for the one-state scalar and the two-state fermion with the two moments alone, and for the two-state fermion after imposing the measured abundance and Pauli exclusion.

%%%%%%%%%%%%%%%%%%%%%%%%%%%%%%%%%%%%%%%%%%%%%%%%%%%%%%%%%%%%%%%%
\begin{figure}[tbp]
  \centering
  \includegraphics[width=.86\linewidth]{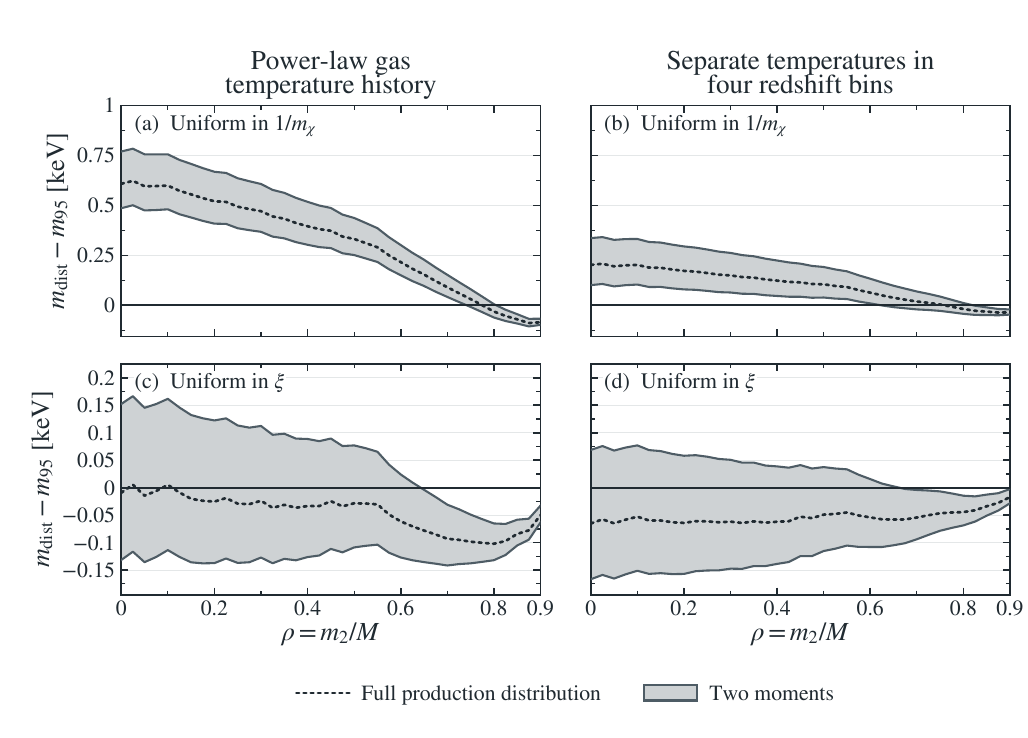}
\caption{Effect of the mass prior for the one-state scalar in three-body decay with the baseline thermal calibration.
The columns use the power-law gas temperature history (left) and separate temperatures in four redshift bins (right).
The priors are uniform in \(1/m_\chi\) (top) and in \(\xi\) (bottom), with \(m_\chi\geq0.75~{\rm keV}\) in both cases.
The gray band spans the distance-matched mass interval relative to \(m_{95}\), and the dotted curve gives \(m_{\rm dist}-m_{95}\) for the full production distribution.
Where the entire band is below zero, \(m_{95}\) exceeds \(m_+\).
The vertical scales differ between rows.}
  \label{fig:three-body-scalar-prior}
\end{figure}
%%%%%%%%%%%%%%%%%%%%%%%%%%%%%%%%%%%%%%%%%%%%%%%%%%%%%%%%%%%%%%%%

In the dilute limit, \(\xi\) scales as \(m_\chi^{-3/4}\), more weakly than linearly with \(1/m_\chi\), so a prior uniform in \(\xi\) shifts prior weight from the low-mass end toward larger masses.
For the one-state scalar in three-body decay, this prior places \(m_{95}\) inside the distance-matched mass interval at every sampled \(\rho\leq0.60\) for both gas temperature histories (Fig.~\ref{fig:three-body-scalar-prior}).
At the two largest sampled mass ratios, \(\rho=0.875\) and \(0.90\), both priors give \(m_{95}>m_+\) for both gas temperature histories with the baseline thermal calibration.

For any mixture of the two normalized mass priors at fixed \(\rho\), 
the posterior probability below \(m_+\) is a weighted average of the probabilities obtained with the two priors separately.
At these two mass ratios, both probabilities are below \(0.05\) for both gas temperature histories with the baseline thermal calibration, so \(m_{95}>m_+\) for every mixture.

At the same two mass ratios, 
the two-state fermion in three-body decay gives \(m_{95}<m_-\) for both mass priors on the same mass range and both gas temperature histories after imposing the measured abundance and Pauli exclusion.

In a separate test at \(\rho=0.50\), we changed both the mass prior and the mass range for the two-state fermion in three-body decay.
For both the thermal relic and the decay model we used a prior uniform in \(\xi\) on \(0\leq\xi\leq0.90\).
For the thermal relic this upper limit corresponds to a minimum mass of \(1/0.90\)~keV.
With the power-law gas temperature history, \(m_{95}\) changes from 3.66~keV to 4.07~keV and crosses both mass thresholds, 3.84 and 3.96~keV.

To predict the flux power for comparison with the HIRES/MIKE measurements, 
we interpolated the linear matter power spectra computed with CLASS in \(\rho\) and in the present root-mean-square momentum per unit mass, \(\sqrt{\mathbb E_{\mathcal P}[r^2]}\).
To test the interpolation of the predicted flux power, we used two alternative choices of weights for the simulated template transfer functions.
The first alternative uses regularized least squares, which penalizes large template weights~\cite{Hooper:2022byl}.
The second alternative uses inverse-distance weighting, 
which gives greater weight to templates whose linear transfer functions are closer to the model on the sampled wavenumber grid.
We used the same interpolation for the thermal relic and the decay model in each fit.

For the one-state scalar in three-body decay, we identify the first sampled mass ratio at which the upper mass threshold \(m_+\) lies below \(m_{95}\) for both gas temperature histories.
The first sampled mass ratio moves from 0.825 in the baseline fit to 0.850 with regularized weights and 0.875 with inverse-distance weights.
Allowing the reionization redshift to vary gives 0.850 with the baseline weights.
With the prior uniform in inverse mass, all these fits retain \(m_->m_{95}\) at every sampled \(\rho\leq0.60\) and \(m_+<m_{95}\) at every sampled ratio from \(0.875\) to \(0.90\).

For the two-state fermion in three-body decay, the ordering at large \(\rho\) depends on the interpolation and on the treatment of perturbations.
The late-time fluid approximation in CLASS closes the evolution equations for the angular moments of the perturbed momentum distribution~\cite{Lesgourgues:2011rh}.
We tested the effect of that approximation by retaining the angular hierarchy throughout the calculation.
With regularized weights and the power-law gas temperature history, 
retaining the angular hierarchy moves \(m_{95}\) across \(m_-\) at \(\rho=0.85\), where the two values nearly coincide.

%%%%%%%%%%%%%%%%%%%%%%%%%%%%%%%%%%%%%%%%%%%%%%%%%%%%%%%%%%%%%%%%

\bibliography{references}

@article{Bode:2000gq,
    author = "Bode, Paul and Ostriker, Jeremiah P. and Turok, Neil",
    title = "{Halo formation in warm dark matter models}",
    eprint = "astro-ph/0010389",
    archivePrefix = "arXiv",
    doi = "10.1086/321541",
    journal = "Astrophys. J.",
    volume = "556",
    pages = "93--107",
    year = "2001"
}

@article{Viel:2005qj,
    author = "Viel, Matteo and Lesgourgues, Julien and Haehnelt, Martin G. and Matarrese, Sabino and Riotto, Antonio",
    title = "{Constraining warm dark matter candidates including sterile neutrinos and light gravitinos with WMAP and the Lyman-alpha forest}",
    eprint = "astro-ph/0501562",
    archivePrefix = "arXiv",
    reportNumber = "DFPD-05-08, LAPTH-1086-05",
    doi = "10.1103/PhysRevD.71.063534",
    journal = "Phys. Rev. D",
    volume = "71",
    pages = "063534",
    year = "2005"
}

@article{Ma:1995ey,
    author = "Ma, Chung-Pei and Bertschinger, Edmund",
    title = "{Cosmological perturbation theory in the synchronous and conformal Newtonian gauges}",
    eprint = "astro-ph/9506072",
    archivePrefix = "arXiv",
    doi = "10.1086/176550",
    journal = "Astrophys. J.",
    volume = "455",
    pages = "7--25",
    year = "1995"
}

@article{Viel:2013fqw,
    author = "Viel, Matteo and Becker, George D. and Bolton, James S. and Haehnelt, Martin G.",
    title = "{Warm dark matter as a solution to the small scale crisis: New constraints from high redshift Lyman-{\ensuremath{\alpha}} forest data}",
    eprint = "1306.2314",
    archivePrefix = "arXiv",
    primaryClass = "astro-ph.CO",
    doi = "10.1103/PhysRevD.88.043502",
    journal = "Phys. Rev. D",
    volume = "88",
    pages = "043502",
    year = "2013"
}

@article{Irsic:2017ixq,
    author = "Ir{\v{s}}i{\v{c}}, Vid and others",
    title = "{New Constraints on the free-streaming of warm dark matter from intermediate and small scale Lyman-$\alpha$ forest data}",
    eprint = "1702.01764",
    archivePrefix = "arXiv",
    primaryClass = "astro-ph.CO",
    doi = "10.1103/PhysRevD.96.023522",
    journal = "Phys. Rev. D",
    volume = "96",
    number = "2",
    pages = "023522",
    year = "2017"
}

@article{Irsic:2023equ,
    author = "Ir{\v{s}}i{\v{c}}, Vid and others",
    title = "{Unveiling dark matter free streaming at the smallest scales with the high redshift Lyman-alpha forest}",
    eprint = "2309.04533",
    archivePrefix = "arXiv",
    primaryClass = "astro-ph.CO",
    doi = "10.1103/PhysRevD.109.043511",
    journal = "Phys. Rev. D",
    volume = "109",
    number = "4",
    pages = "043511",
    year = "2024"
}

@article{Merle:2015oja,
    author = "Merle, Alexander and Totzauer, Maximilian",
    title = "{keV Sterile Neutrino Dark Matter from Singlet Scalar Decays: Basic Concepts and Subtle Features}",
    eprint = "1502.01011",
    archivePrefix = "arXiv",
    primaryClass = "hep-ph",
    reportNumber = "MPP-2015-13",
    doi = "10.1088/1475-7516/2015/06/011",
    journal = "{JCAP}",
    volume = "06",
    pages = "011",
    year = "2015"
}

@article{Heeck:2017xbu,
    author = "Heeck, Julian and Teresi, Daniele",
    title = "{Cold keV dark matter from decays and scatterings}",
    eprint = "1706.09909",
    archivePrefix = "arXiv",
    primaryClass = "hep-ph",
    reportNumber = "ULB-TH-17-09",
    doi = "10.1103/PhysRevD.96.035018",
    journal = "Phys. Rev. D",
    volume = "96",
    number = "3",
    pages = "035018",
    year = "2017"
}

@article{DEramo:2020gpr,
    author = "D'Eramo, Francesco and Lenoci, Alessandro",
    title = "{Lower mass bounds on FIMP dark matter produced via freeze-in}",
    eprint = "2012.01446",
    archivePrefix = "arXiv",
    primaryClass = "hep-ph",
    reportNumber = "DESY 20-219, DESY-20-219",
    doi = "10.1088/1475-7516/2021/10/045",
    journal = "{JCAP}",
    volume = "10",
    pages = "045",
    year = "2021"
}

@article{Decant:2021mhj,
    author = "Decant, Quentin and Heisig, Jan and Hooper, Deanna C. and Lopez-Honorez, Laura",
    title = "{Lyman-{\ensuremath{\alpha}} constraints on freeze-in and superWIMPs}",
    eprint = "2111.09321",
    archivePrefix = "arXiv",
    primaryClass = "astro-ph.CO",
    reportNumber = "ULB-TH/21-20, TTK-21-46, HIP-2021-38/TH",
    doi = "10.1088/1475-7516/2022/03/041",
    journal = "{JCAP}",
    volume = "03",
    pages = "041",
    year = "2022"
}

@article{Ballesteros:2020adh,
    author = "Ballesteros, Guillermo and Garcia, Marcos A. G. and Pierre, Mathias",
    title = "{How warm are non-thermal relics? Lyman-$\alpha$ bounds on out-of-equilibrium dark matter}",
    eprint = "2011.13458",
    archivePrefix = "arXiv",
    primaryClass = "hep-ph",
    reportNumber = "IFT-UAM/CSIC-20-135",
    doi = "10.1088/1475-7516/2021/03/101",
    journal = "{JCAP}",
    volume = "03",
    pages = "101",
    year = "2021"
}

@article{DEramo:2025jsb,
    author = "D'Eramo, Francesco and Lenoci, Alessandro and Dekker, Ariane",
    title = "{Dark matter freeze-in and small-scale observables: Novel mass bounds and viable particle candidates}",
    eprint = "2506.13864",
    archivePrefix = "arXiv",
    primaryClass = "hep-ph",
    doi = "10.1103/j62q-cvkr",
    journal = "Phys. Rev. D",
    volume = "112",
    number = "11",
    pages = "116008",
    year = "2025"
}

@article{Liu:2024pjg,
    author = "Liu, Rayne and Hu, Wayne and Xiao, Huangyu",
    title = "{Warm and fuzzy dark matter: Free streaming of wave dark matter}",
    eprint = "2406.12970",
    archivePrefix = "arXiv",
    primaryClass = "hep-ph",
    reportNumber = "FERMILAB-PUB-24-0296-T",
    doi = "10.1103/PhysRevD.111.023535",
    journal = "Phys. Rev. D",
    volume = "111",
    number = "2",
    pages = "023535",
    year = "2025"
}

@article{GPP1998,
    author = "Gulja{\v s}, B. and Pearce, C. E. M. and Pe{\v c}ari{\'c}, J.",
    title = "{Jensen's inequality for distributions possessing higher moments, with application to sharp bounds for Laplace-Stieltjes transforms}",
    journal = "J. Austral. Math. Soc. Ser. B",
    volume = "40",
    number = "1",
    pages = "80--85",
    year = "1998",
    doi = "10.1017/S0334270000012388"
}

@article{Jensen1906,
    author = {Jensen, J. L. W. V.},
    title = "{Sur les fonctions convexes et les in\'egalit\'es entre les valeurs moyennes}",
    doi = "10.1007/BF02418571",
    journal = "Acta Math.",
    volume = "30",
    pages = "175--193",
    year = "1906"
}

@article{Tremaine:1979we,
    author = "Tremaine, S. and Gunn, J. E.",
    editor = "Srednicki, M. A.",
    title = "{Dynamical Role of Light Neutral Leptons in Cosmology}",
    doi = "10.1103/PhysRevLett.42.407",
    journal = "Phys. Rev. Lett.",
    volume = "42",
    pages = "407--410",
    year = "1979"
}

@article{Planck:2018vyg,
    author = "Aghanim, N. and others",
    collaboration = "Planck",
    title = "{Planck 2018 results. VI. Cosmological parameters}",
    eprint = "1807.06209",
    archivePrefix = "arXiv",
    primaryClass = "astro-ph.CO",
    doi = "10.1051/0004-6361/201833910",
    journal = "Astron. Astrophys.",
    volume = "641",
    pages = "A6",
    year = "2020",
    note = "[Erratum: Astron.Astrophys. 652, C4 (2021)]"
}

@article{Fixsen:2009ug,
    author = "Fixsen, D. J.",
    title = "{The Temperature of the Cosmic Microwave Background}",
    eprint = "0911.1955",
    archivePrefix = "arXiv",
    primaryClass = "astro-ph.CO",
    doi = "10.1088/0004-637X/707/2/916",
    journal = "Astrophys. J.",
    volume = "707",
    pages = "916--920",
    year = "2009"
}

@article{Lesgourgues:2011rh,
    author = "Lesgourgues, Julien and Tram, Thomas",
    title = "{The Cosmic Linear Anisotropy Solving System (CLASS) IV: efficient implementation of non-cold relics}",
    eprint = "1104.2935",
    archivePrefix = "arXiv",
    primaryClass = "astro-ph.CO",
    reportNumber = "CERN-PH-TH-2011-084, LAPTH-012-11",
    doi = "10.1088/1475-7516/2011/09/032",
    journal = "{JCAP}",
    volume = "09",
    pages = "032",
    year = "2011"
}

@article{Blas:2011rf,
    author = "Blas, Diego and Lesgourgues, Julien and Tram, Thomas",
    title = "{The Cosmic Linear Anisotropy Solving System (CLASS) II: Approximation schemes}",
    eprint = "1104.2933",
    archivePrefix = "arXiv",
    primaryClass = "astro-ph.CO",
    reportNumber = "CERN-PH-TH-2011-082, LAPTH-010-11",
    doi = "10.1088/1475-7516/2011/07/034",
    journal = "{JCAP}",
    volume = "07",
    pages = "034",
    year = "2011"
}

@article{Du:2021jcj,
    author = "Du, Yong and Huang, Fei and Li, Hao-Lin and Li, Yuan-Zhen and Yu, Jiang-Hao",
    title = "{Revisiting dark matter freeze-in and freeze-out through phase-space distribution}",
    eprint = "2111.01267",
    archivePrefix = "arXiv",
    primaryClass = "hep-ph",
    reportNumber = "UCI-HEP-TR-2021-27, Report-no: UCI-HEP-TR-2021-27",
    doi = "10.1088/1475-7516/2022/04/012",
    journal = "{JCAP}",
    volume = "04",
    number = "04",
    pages = "012",
    year = "2022"
}

@article{Konig:2016dzg,
    author = {K{\"o}nig, Johannes and Merle, Alexander and Totzauer, Maximilian},
    title = "{keV Sterile Neutrino Dark Matter from Singlet Scalar Decays: The Most General Case}",
    eprint = "1609.01289",
    archivePrefix = "arXiv",
    primaryClass = "hep-ph",
    reportNumber = "MPP-2016-263",
    doi = "10.1088/1475-7516/2016/11/038",
    journal = "{JCAP}",
    volume = "11",
    pages = "038",
    year = "2016"
}

@article{Hooper:2022byl,
    author = {Hooper, Deanna C. and Sch{\"o}neberg, Nils and Murgia, Riccardo and Archidiacono, Maria and Lesgourgues, Julien and Viel, Matteo},
    title = "{One likelihood to bind them all: Lyman-{\ensuremath{\alpha}} constraints on non-standard dark matter}",
    eprint = "2206.08188",
    archivePrefix = "arXiv",
    primaryClass = "astro-ph.CO",
    reportNumber = "HIP-2022-17/TH, TTK-22-21",
    doi = "10.1088/1475-7516/2022/10/032",
    journal = "{JCAP}",
    volume = "10",
    pages = "032",
    year = "2022"
}

@article{Harris:2020xlr,
    author = "Harris, Charles R. and others",
    title = "{Array programming with NumPy}",
    eprint = "2006.10256",
    archivePrefix = "arXiv",
    primaryClass = "cs.MS",
    doi = "10.1038/s41586-020-2649-2",
    journal = "Nature",
    volume = "585",
    number = "7825",
    pages = "357--362",
    year = "2020"
}

@article{Virtanen:2019joe,
    author = "Virtanen, Pauli and others",
    title = "{SciPy 1.0--Fundamental Algorithms for Scientific Computing in Python}",
    eprint = "1907.10121",
    archivePrefix = "arXiv",
    primaryClass = "cs.MS",
    doi = "10.1038/s41592-019-0686-2",
    journal = "Nature Meth.",
    volume = "17",
    pages = "261",
    year = "2020"
}

@article{Hunter:2007ouj,
    author = "Hunter, John D.",
    title = "{Matplotlib: A 2D Graphics Environment}",
    doi = "10.1109/MCSE.2007.55",
    journal = "Comput. Sci. Eng.",
    volume = "9",
    number = "3",
    pages = "90--95",
    year = "2007"
}

@article{McFadden1975,
    author = "McFadden, Daniel",
    title = "{Tchebyscheff bounds for the space of agent characteristics}",
    journal = "J. Math. Econ.",
    volume = "2",
    number = "2",
    pages = "225--242",
    year = "1975",
    doi = "10.1016/0304-4068(75)90025-7"
}

@article{Cerf1994,
    author = "Cerf, Rapha{\"e}l and Mariconda, Carlo",
    title = "{Oriented measures with continuous densities and the bang-bang principle}",
    journal = "J. Funct. Anal.",
    volume = "126",
    number = "2",
    pages = "476--505",
    year = "1994",
    doi = "10.1006/jfan.1994.1155"
}

@article{Saikawa:2018rcs,
    author = "Saikawa, Ken'ichi and Shirai, Satoshi",
    title = "{Primordial gravitational waves, precisely: The role of thermodynamics in the Standard Model}",
    eprint = "1803.01038",
    archivePrefix = "arXiv",
    primaryClass = "hep-ph",
    reportNumber = "IPMU18-0037, MPP-2018-19",
    doi = "10.1088/1475-7516/2018/05/035",
    journal = "{JCAP}",
    volume = "05",
    pages = "035",
    year = "2018"
}

@article{Strumia:2022qvj,
    author = "Strumia, Alessandro",
    title = "{Dark Matter from freeze-in and its inhomogeneities}",
    eprint = "2211.08359",
    archivePrefix = "arXiv",
    primaryClass = "hep-ph",
    doi = "10.1007/JHEP03(2023)042",
    journal = "{JHEP}",
    volume = "03",
    pages = "042",
    year = "2023"
}

@article{Esposito:2022plo,
    author = "Esposito, Matteo and Ir{\v{s}}i{\v{c}}, Vid and Costanzi, Matteo and Borgani, Stefano and Saro, Alexandro and Viel, Matteo",
    title = "{Weighing cosmic structures with clusters of galaxies and the intergalactic medium}",
    eprint = "2202.00974",
    archivePrefix = "arXiv",
    primaryClass = "astro-ph.CO",
    doi = "10.1093/mnras/stac1825",
    journal = "Mon. Not. Roy. Astron. Soc.",
    volume = "515",
    number = "1",
    pages = "857--870",
    year = "2022"
}

@article{Dienes:2021cxp,
    author = "Dienes, Keith R. and Huang, Fei and Kost, Jeff and Thomas, Brooks and Yu, Hai-Bo",
    title = "{Evaluating Lyman-{\ensuremath{\alpha}} constraints for general dark-matter velocity distributions: Multiple scales and cautionary tales}",
    eprint = "2112.09105",
    archivePrefix = "arXiv",
    primaryClass = "astro-ph.CO",
    reportNumber = "UCI-HEP-TR-2021-31",
    doi = "10.1103/PhysRevD.106.123521",
    journal = "Phys. Rev. D",
    volume = "106",
    number = "12",
    pages = "123521",
    year = "2022"
}

@article{Murgia:2018now,
    author = "Murgia, Riccardo and Ir{\v{s}}i{\v{c}}, Vid and Viel, Matteo",
    title = "{Novel constraints on noncold, nonthermal dark matter from Lyman- {\ensuremath{\alpha}} forest data}",
    eprint = "1806.08371",
    archivePrefix = "arXiv",
    primaryClass = "astro-ph.CO",
    doi = "10.1103/PhysRevD.98.083540",
    journal = "Phys. Rev. D",
    volume = "98",
    number = "8",
    pages = "083540",
    year = "2018"
}

@article{Carena:2021bqm,
    author = "Carena, Marcela and Coyle, Nina M. and Li, Ying-Ying and McDermott, Samuel D. and Tsai, Yuhsin",
    title = "{Cosmologically degenerate fermions}",
    eprint = "2108.02785",
    archivePrefix = "arXiv",
    primaryClass = "hep-ph",
    reportNumber = "FERMILAB-PUB-21-325-T",
    doi = "10.1103/PhysRevD.106.083016",
    journal = "Phys. Rev. D",
    volume = "106",
    number = "8",
    pages = "083016",
    year = "2022"
}

@article{Hui:1997dp,
    author = "Hui, Lam and Gnedin, Nickolay Y.",
    title = "{Equation of state of the photoionized intergalactic medium}",
    eprint = "astro-ph/9612232",
    archivePrefix = "arXiv",
    reportNumber = "FERMILAB-PUB-97-027-A",
    doi = "10.1093/mnras/292.1.27",
    journal = "Mon. Not. Roy. Astron. Soc.",
    volume = "292",
    pages = "27",
    year = "1997"
}

@article{Mangano:2005cc,
    author = "Mangano, Gianpiero and Miele, Gennaro and Pastor, Sergio and Pinto, Teguayco and Pisanti, Ofelia and Serpico, Pasquale D.",
    title = "{Relic neutrino decoupling including flavor oscillations}",
    eprint = "hep-ph/0506164",
    archivePrefix = "arXiv",
    reportNumber = "DSF-16-2005, IFIC-05-17, MPP-2005-36",
    doi = "10.1016/j.nuclphysb.2005.09.041",
    journal = "Nucl. Phys. B",
    volume = "729",
    pages = "221--234",
    year = "2005"
}

@article{Hall:2009bx,
    author = "Hall, Lawrence J. and Jedamzik, Karsten and March-Russell, John and West, Stephen M.",
    title = "{Freeze-In Production of FIMP Dark Matter}",
    eprint = "0911.1120",
    archivePrefix = "arXiv",
    primaryClass = "hep-ph",
    reportNumber = "OUTP-09-18-P, UCB-PTH-09-32",
    doi = "10.1007/JHEP03(2010)080",
    journal = "JHEP",
    volume = "03",
    pages = "080",
    year = "2010"
}
\end{document}